\documentclass[11pt,superscriptaddress,floatfix]{article}
\pdfoutput=1
\usepackage{epsfig}
\usepackage{epstopdf}
\usepackage{graphicx,psfrag}
\usepackage{amssymb}

\usepackage[utf8]{inputenc}
\usepackage{jheppub}
\usepackage{color}
\usepackage{amsmath}
\usepackage{hyperref}
\usepackage{setspace}
\usepackage{array}
\usepackage{tikz}
\usepackage{tkz-euclide}
\usepackage{amsthm}
\usepackage{bbold}
\usepackage[utf8]{inputenc}

\newcommand{\Tr}{\rm Tr}
\newcommand{\eref}[1]{(\ref{#1})}
\newcommand{\nn}{\nonumber}

\newcommand{\be}{\begin{eqnarray}}
\newcommand{\ee}{\end{eqnarray}}

\newcommand{\bmat}{\left ( \begin{array}{cc} }
\newcommand{\emat}{\end{array} \right ) }

\def\Tr{\textrm{Tr}}

\newcommand{\beq}{\begin{equation}}
\newcommand{\beqs}{\begin{equation*}}
\newcommand{\eeq}{\end{equation}}
\newcommand{\eeqs}{\end{equation*}}

\begin{document}

\title{Large $N$ expansion of the moments and free energy of Sachdev-Ye-Kitaev model, and the enumeration
  of intersection graphs}

\author[a]{Yiyang Jia}
\author[a]{and Jacobus J. M. Verbaarschot}

\affiliation[a]{Department of Physics and Astronomy, Stony Brook University, Stony Brook, New York 11794, USA}

\emailAdd{yiyang.jia@stonybrook.edu}
\emailAdd{jacobus.verbaarschot@stonybrook.edu}

\date{\today}
\abstract{
  In this paper we explain the relation between the free energy of the SYK model for $N$ Majorana fermions with a random $q$-body interaction
  and the moments of its
  spectral density.
  The high temperature expansion of the free energy gives the cumulants of the spectral
  density. Using that the cumulants are extensive we find the $p$ dependence of the $1/N^2$ correction 
  of the $2p$-th moments obtained in \cite{GG-J-V}. Conversely, the $1/N^2$ corrections to the
  moments give the correction (even  $q$) to the $\beta^6$ coefficient of the high temperature expansion of the free energy for
  arbitrary $q$. Our result agrees with the $1/q^3$ correction
  obtained by Tarnopolsky
  using a mean field expansion. 
  These considerations also lead to a more powerful method for solving the moment problem and  intersection-graph enumeration problems. We take advantage of this and push the moment calculation to $1/N^3$ order and find
  surprisingly simple enumeration identities for intersection graphs.
  The $1/N^3$ corrections to the moments, give  corrections to
 the  $\beta^8$ coefficient (for even $q$) of the high temperature expansion of the  free energy which have not been calculated before.
  Results for odd $q$, where the SYK ``Hamiltonian'' is the supercharge of a supersymmetric
  theory are discussed as well.
  }

\maketitle\flushbottom
\newpage
\section{Introduction}

The statistical fluctuations of nuclear levels have been successfully described
by the Gaussian Orthogonal Ensemble (GOE). However, the average spectral density of the GOE
is a semicircle which is very different from the Bethe formula \cite{bethe1936}.
In agreement with experimental observations \cite{Egidy:2005aw}, this formula
predicts an $\exp (c\sqrt {E-E_0})$ dependence (with $E_0$ the ground state energy) on the excitation energy $E$.
Moreover, the nuclear interaction is
mostly a two-body interaction while for the GOE all the many-body states interact.
 To address these shortcomings, French and co-workers \cite{french1970,french1971,bohigas1971,bohigas1971a} introduced
the two-body random ensemble which is now known as the four-body
complex SYK model. One of the first results for this ensemble was that the
level density is a Gaussian \cite{bohigas1971}  which is closer to the
expectation of realistic many-body systems
 than the
semicircular behavior. However, in the nuclear physics community it was
not realized that this model actually reproduces the Bethe formula \cite{garcia2017}. One of the
reasons for missing this opportunity  was the custom
\cite{brody1981,kota2001,gomez2011,kota2011a,kota2014}
to study this model as the sum of a two-body and a four-body interaction (in
the nuclear physics convention,  a one-body and a two-body interaction).
The reason is that the nuclear interaction was seen as a residual
random four-body interaction on top of a mean field which can be represented as a two-body interaction.
Note
that the four-body interaction is a irrelevant  term with regard to the two-body interaction \cite{garcia2017a,Nosaka:2018iat,Garcia-Garcia:2018ruf}.
It was also understood early on that the two-body random ensemble still has
the level correlations of the Gaussian Orthogonal Ensemble \cite{bohigas1971a}. However, it
was realized that the model was not ergodic \cite{Bohigas:1974zz,benet2001,Altland:2017eao,Garcia-Garcia:2018ruf} in the sense that
ensemble average of a spectral correlator is not equal to the spectral
average of a spectral correlator, the latter given by the result of   the Wigner-Dyson
ensembles up to much larger distances. For more discussions of the two-body random
ensemble in nuclear and many-body physics we refer to
\cite{brody1981,benet2001,benet2003,borgonovi2016}.

In condensed matter, the model was introduced independently as a  random quantum spin
model \cite{sachdev1993}
where a mean field is not natural. In this context, Sachdev and
co-workers discovered \cite{sachdev00} a remarkable property of this model, namely that
its zero temperature entropy is extensive which then was identified as the
black hole entropy \cite{Sachdev:2010um}.
This property is
directly related to its exponentially large (with respect to system size) level density starting with the ground state region,
which means this model is a non-Fermi liquid.  The states are  characterized  by highly  entangled states
\cite{Huang:2017nox} which are very
different from the particle-hole excitations of a Fermi liquid.

In the past two years the interest in this model was rekindled because
it is possibly dual to 1+1 dimensional
gravity \cite{kitaev2015,maldacena2016,jensen2016,polchinski2016,polchinski2016a,Jevicki:2016ito,Jevicki:2016bwu,garcia2016,witten2016,garcia2017,Das:2017pif,Das:2017wae,cotler:2017jue,cotler2016,Krishnan:2016bvg,Turiaci:2017zwd,Klebanov:2016xxf,Stanford:2017thb,gross2017,Gross:2017aos,Bagrets:2017pwq,Cai:2017vyk}.
The properties that made this model attractive as a model of a compound
nucleus are exactly those which are required for the existence of a black-hole dual: it is maximally chaotic \cite{maldacena2015}
with spectral correlations given by
random matrix theory \cite{you2016,garcia2016,Garcia-Garcia:2018ruf}, it has a level density given by the Bethe formula
which also implies that the specific heat is linear in temperature for low
temperature, the  zero temperature entropy is extensive showing that the low-lying
states of the model are strongly entangled.
What is particularly important in this context
is the existence of a conformal limit \cite{kitaev2015,maldacena2016} where
the action of the SYK model reduces to the Schwarzian action
\cite{maldacena2016,Bagrets:2016cdf,altland2017,Stanford:2017thb}.

Much of the recent progress on the SYK model was made possible because
of the path integral formulation
\cite{sachdev1993,kitaev2015,maldacena2016},
from which one can derive an exact
result for large
$N$ limit after averaging over the random interaction. This was not possible for
a formulation that started from the generating function for the resolvent
\cite{verbaarschot1984}. Although it was straightforward to average over the randomness,
it resulted in a complicated theory that was not amenable to taking a large $N$ limit.
The disadvantage of the path integral which at the same time is its
strength is that it provides access to the
Green's function rather than the level density.  Since our main interest has
been in the level density and the level correlations
\cite{garcia2016,garcia2017,Garcia-Garcia:2018ruf,GG-J-V,garcia2017a}
of the SYK model, we have
used the moment method \cite{mon1975}
which also proved effective in the early applications to
nuclear physics. Two limiting cases for  $N$ fermions
fermions with a random $q$-body interaction
were easily recognized: $ q^2\gg N$
and $ q^2\ll N$. In the first case  \cite{benet2001,Liu:2016rdi} the SYK model is in the universality class
of the Wigner-Dyson ensembles with a semicircular spectral density. In the
second case, the spectral density is a Gaussian \cite{bohigas1971}. This
suggests the existence of a double scaling limit which converges to a spectral
density in between a semicircle and a Gaussian. Indeed this happens when
$q \sim \sqrt N$ for $N\to \infty$ \cite{erdos2014}. This scaling limit which reveals itself in
the path integral formulation of the SYK model  was not noted before
in the nuclear physics literature either. In the moment method, this limit arises
naturally in Wick contractions when treating all intersections as independent which gives approximate moments that are exact to order $1/N$
\cite{erdos2014,garcia2016,cotler2016,garcia2017}.
Remarkably, these moments turn
out to give the spectral density of the weight function of the Q-Hermite polynomials with a nontrivial double scaling limit
\cite{erdos2014,cotler2016,garcia2017,Feng:2018zsx}. In this paper, when we speak of ``corrections'', we typically mean corrections to the Q-Hermite result.

The $1/N^2$ corrections to all moments can be calculated analytically
\cite{GG-J-V},
with a result that has as a simple geometric interpretation. The $p$-dependence
of the $2p$-th moment also turns out to be relatively simple. One of the main  goals
of  this paper is  to
explain this $p$-dependence of the moments.
Since the high temperature expansion of
the free energy is the  cumulant expansion of the spectral density, the extensivity
of the free energy puts strong constraints on the moments. In fact, we will show
that the $p$-dependence of the moments follows almost entirely from this condition, and
that it is determined by a few low-order moments only.
Secondly, we study the way  large $N$ corrections to the moments contribute to the free energy. We already know that the
Q-Hermite moments give the free energy for all temperatures to order $1/q^2$ \cite{garcia2017,Tarnopolsky}.
In this
paper we will show that corrections to the $\beta^6$ high-temperature coefficient of the free energy (for even $q$) follow from the
$1/N^2$ corrections to moments. The result is valid for all $q$ and it gives the
$1/q^3$ corrections to the free energy.
Thirdly,  using the relationship between
the free energy and the moments, we  obtain the $1/N^3$ corrections to the
moments which are responsible for the correction to the  $\beta^8$ coefficient in the high temperature expansion of the free energy (for even $q$). We stress that our high temperature expansion results are valid for any finite $q$, but because some of the results are organized in powers of $1/q$, we sometimes use language such as ``large $q$ expansion'' and ``large $q$ corrections''.
 Using these results we find new enumerative
identities for intersection graphs.
Results for the supersymmetric SYK
model \cite{fu2017,li2017,kanazawa2017} which can be derived in a similar way are
also given in this paper.

This paper is organized as follows.
In section 2, we give a brief review of the SYK model including
the moment method. The relation between moments and the free energy is
discussed in section 3. In this section we show that at a given order in $1/N$ the $p$-dependence of
the moments follows from a few low-order moments.
We also show that large $N$ corrections to the moments give large $q$ corrections
to the free energy. Results are obtained for both  even $q$ and odd $q$.
As a new result we obtain the $1/N^3$ corrections to the moments and $\beta^4$  corrections to the free energy, which  give the $1/q^4$ corrections to the free energy.
In section 4, we use the results for the moments to
derive new enumerative identities for intersection graphs. Some technical
results are deferred to the appendices.
In appendix
\ref{append:tarnopolsky}, we evaluate the high temperature expansion to the free energy
 from the results of Tarnopolsky \cite{Tarnopolsky}. We obtain the high temperature
expansion of Tarnopolsky's result to all orders and show that it is a convergent series with no
singularities on the positive real axis. In appendix \ref{appen:freeEnergyNcube} we write down the high temperature expansion of the free energies to order $1/N^3$. Finally,  we derive in appendix \ref{sec:consistency}  the
$1/N^3$ corrections for $ q=1$ and $q=2$ models with an independent method.

 \section{Review of SYK model and moment method}

 In this section we introduce the Sachdev-Ye-Kitaev (SYK) model and the moment method.
\subsection{The SYK Hamiltonian}
The  SYK model is a system of $N$ Majorana particles with the $q$-body interaction represented by
\be
H = \sum J_\alpha \Gamma_\alpha
\label{syk}
\ee
with 
\begin{equation}
\Gamma_\alpha=(i)^{{q(q-1)}/{2}}\gamma_{i_1}\gamma_{i_2}\cdots\gamma_{i_q},
\end{equation}
and $\gamma_\alpha$ are the Euclidean gamma matrices with the anticommutation relation:
\be\label{eqn:gammAntiComm}
\{\gamma_k,\gamma_l\} = 2\delta_{kl}.
\ee
The factors of $i$ in the definition of $\Gamma_\alpha$ have been included
so that $H$ is Hermitian.
The sum is over all ${N \choose q}$ $q$-particle index denoted by the collective index 
$\alpha=\{i_1,i_2,\ldots,i_q\}$, with $1\leq i_1<i_2<\cdots<i_q\leq N$.  The couplings $J_\alpha$ are Gaussian distributed:
\begin{equation}\label{GaussianDistri}
P(J_\alpha)=\sqrt{\frac{2^{q-1} N^{q-1}}{(q-1)!\pi J^2}}\exp\left(-\frac {2^{q-1}N^{q-1}J_\alpha^2}{(q-1)!J^2}\right),
\end{equation}
where the parameter $J$ sets a physical scale.\footnote{A factor of $2^q$ has been included in the variance so that our results coincide with 
the  Majorana convention $\{\gamma_k, \gamma_l\}=\delta_{kl}$.}
For even $q$ the Hamiltonian $\cal H$ of the SYK model is simply given by
\be
   {\cal H} = H.
 \ee  
 With the large $N$ scaling of the variance in \eref{GaussianDistri} this Hamiltonian
 has a negative-energy ground state energy that is proportional to $N$ for large $N$.

For odd $q$ the operator $H$ is still a well-defined Hermitian operator, but because it has a fermionic grading, it
is not a Hamiltonian but rather the supercharge of a supersymmetric theory with the Hamiltonian \cite{fu2017}:
\be
{\cal H} = H^2.
\ee
This Hamiltonian is positive definite with a ground state energy approaching
zero in the thermodynamic limit with the scaling of the variance as in
\eref{GaussianDistri}.

 \subsubsection{Moments and spectral density}
 One of the reasons for which we are interested in moments is to study the spectral density $\rho(E)$ of the SYK model:\footnote{As we will see, the spectral density will not play an important role in the higher-order free energy calculation, which is the main subject of this paper. Nevertheless, the spectral density
   provides a natural introduction of the moments.}
 \begin{equation}
\rho(\lambda):=\left\langle \sum\limits_{k=1}^{2^{\lfloor \frac{N}{2} \rfloor}}\delta(\lambda-\lambda_{k}) \right \rangle,
\end{equation}  
where $\langle\cdots\rangle$ denotes the Gaussian average using the distribution \eref{GaussianDistri}.
By a Fourier transform, we can express $\rho(\lambda)$ as 
\begin{equation}\label{eqn:SpectDenMom}
\begin{split}
\rho(\lambda)&=\frac{1}{2\pi}\int_{-\infty}^{\infty}\text{d}te^{-iEt}\left\langle \text{Tr} e^{iHt}\right \rangle\\
&= 2^{N/2} \frac{1}{2\pi}\int_{-\infty}^{\infty}\text{d}te^{-i\lambda t}\sum\limits_{k=0}^{\infty}\frac{(it)^{k}}{(k)!}M_{k}
\\
&= 2^{N/2} \frac{1}{2\pi}\int_{-\infty}^{\infty}\text{d}te^{-i\lambda t}\sum\limits_{p=0}^{\infty}\frac{(it)^{2p}}{(2p)!}M_{2p},
\end{split}
\end{equation}
where we have defined the $k$-th moment $M_k$ to be
\begin{equation}
M_k:=\left\langle \text{Tr} H^k\right \rangle/2^{N/2}= 2^{-N/2}\int d\lambda \rho(\lambda) \lambda^k.
\end{equation}
The third equality of eq. \eref{eqn:SpectDenMom} used the fact that $M_k=0$ when $k$ is odd, due to the $J_\alpha\to -J_\alpha$ symmetry of the distribution \eref{GaussianDistri}.
Therefore, we may focus on the calculation of $M_{2p}$. Notice that since $\langle\cdots\rangle$ is an Gaussian integration over $J_\alpha$, $M_{2p}$ is given by a sum over all possible Wick contractions among $2p$ $\Gamma$'s.   By convention, we have factored out the Hilbert space dimensionality $2^{N/2}$, but $M_{2p}$ still grows like $N^p$  as $N\to \infty$, as can be seen from the $N$ dependence of $M_2$:\be
 \label{eqn:m2}
 M_2=\sigma^2 = \binom{N}{q}\frac{(q-1)!J^2}{2^q N^{q-1}}.
 \ee
Hence, to formulate a useful large $N$ expansion for moments, we consider instead the scaled moments:
\begin{equation}
\tilde M_{2p} := \frac{M_{2p}}{M_2^p}\sim O(1).
\end{equation}

We distinguish the moments of the SYK operator $H$, which will be
denoted by $M_{2p}$ and the moments of the Hamiltonian ${\cal H}$ which
will  be denoted by $\mu_p$. So we have that  
\be
  \mu_{p} &=& M_p, \qquad {\text{for even $q$} },\label{eqn:EvenqMandMu}\\  
  \mu_{p} &=& M_{2p}, \qquad {\text{for odd $q$}}.  \label{eqn:OddqMandMu}
  \ee
  
\section{Free energy, cumulants and high temperature expansions}
\subsection{Moments and cumulants }
The partition function is given by
\be
Z = {\rm Tr} e^{-\beta {\cal H}} \equiv e^{-\beta F}
\ee
with the free energy denoted by $F$.
In terms of high temperature expansion of the free energy, we have 
\begin{equation}\label{eqn:HighTempCum}
-\beta F -\frac{N}{2}\log 2 = \sum_{n=1}^{\infty}\frac{\kappa_n}{n!}(-\beta)^n.
\end{equation}
The quantity $\kappa_n$ is called the $n$-th cumulant. Note that the summation starts from $n=1$, this is because the energy is finite, and at infinite temperature we expect only the entropy to contribute to $\beta F$. Since free energy is extensive, we obtain 
\begin{equation}\label{eqn:CumAsym}
\kappa_n\sim O(N)
\end{equation}
for all $n$. We will see later that this has important
consequences for the $N$-dependence of the moments.

Alternatively, the partition function can also be expressed in the moments
$\mu_n$ of ${\cal H}$:
\be
Z = \sum_{n=0}^{\infty} \frac {\mu_n}{n!} (-\beta)^n.
\ee
One may ask why we do not consider the ``free energy'' $e^{-\beta F}:=\langle\Tr (e^{-\beta H})\rangle$ also for odd $q$ so that we can have a uniform treatment of the moments for all values of $q$, instead of two separate cases eqs. \eqref{eqn:EvenqMandMu} and \eqref{eqn:OddqMandMu}. The problem with this ``free energy'' is that it is not extensive. Extensivity will turn out to be vital for the application of our method.\\

The relation between $\mu_n$ and $\kappa_n$ is  well studied \cite{cumulants}. Consider the following partition of an integer $n$:
\begin{equation}
n = \underbrace{k_1+\cdots k_1}_{m_1 \text{ times}}+\underbrace{k_2+\cdots+k_2}_{m_2 \text{ times}}+\cdots+\underbrace{k_l+\cdots k_l}_{m_l \text{ times}}=\sum_{i=1}^l m_i k_i,
\end{equation}
where by convention we demand the ordering $k_1\geq k_2\geq\cdots \geq k_l$,\footnote{Alternatively we can say, for example, $3=2+1$ and $3=1+2$ are the same partition of $3$.} then we have 
\begin{align}
\mu_n &= \sum_{P_n}\left( \frac{n!}{\prod_{i=1}^l  m_i!(k_i!)^{m_i}} \prod_{j=1}^l (\kappa_{k_j})^{m_j}\right),\label{eqn:CumToMomConvert}\\
\kappa_n &= \sum_{P_n}\left((-1)^{\sum_i m_i -1}\left(\sum_{i=1}^l m_i -1\right)! \ \frac{n!}{\prod_{i=1}^l  m_i!(k_i!)^{m_i}} \prod_{j=1}^l (\mu_{k_j})^{m_j}\right),\label{eqn:MomToCumConvert}
\end{align}
where $\sum_{P_n}$ denotes sum over all partitions of $n$. 
As examples, we list some low-order relations:
\begin{equation}
\begin{split}
&\mu_1=\kappa_1,\quad \mu_2=\kappa_2+\kappa_1^2,\quad \mu_3=\kappa_3+3\kappa_2\kappa_1+\kappa_1^3;\\
&\kappa_1=\mu_1,\quad \kappa_2 = \mu_2-\mu_1^2,\quad \kappa_3 = \mu_3-3\mu_2\mu_1+2\mu_1^2. 
\end{split}
\end{equation}
We remark that since moments are computed by contracting $\Gamma$ matrices, all the $N$-dependence comes from counting the subscripts  of those  $\Gamma$ matrices. This implies the $\mu_n$ must be rational functions of $N$, and then eq. \eqref{eqn:MomToCumConvert} tells us the cumulants $\kappa_n$ must also be rational functions of $N$. Hence there can be no factors such as $\log N$ or $\sqrt N$ in the large $N$ expansion of moments or cumulants.
For even $q$ all odd cumulants vanish, but they enter in the calculations for odd $q$.

\subsection{Even $q$ case}

\subsubsection{$N$ dependence of cumulants and moments}
\label{sec:3.2.1}

As discussed, for even $q$ we have $M_k=\mu_k$ and hence $\mu_k=0$ when $k$ is an odd number. It follows from  eq. \eref{eqn:MomToCumConvert} that $\kappa_k=0$ when $k$ is odd. We are interested in the scaled moments $\tilde M_{2p}$, so let us also define the scaled cumulants to be 
\begin{equation}
\tilde \kappa_{2k} := \frac{\kappa_{2k}}{\kappa_2^k} =  \frac{\kappa_{2k}}{M_2^k}=  \frac{\kappa_{2k}}{\sigma^{2k}}.
\end{equation}
According to eq. \eqref{eqn:m2}, $M_{2}=\sigma^2= \binom{N}{q} \frac{J^2(q-1)!}{2^qN^{q-1}}$ which is  $\sim N$, together with the free energy extensivity eq. \eqref{eqn:CumAsym},  we deduce

\begin{equation}
  \label{eqn:scaledCumuOrder}
\tilde \kappa_{2k}\sim O(N^{-k+1}).
\end{equation}
We also have the trivial identity
\begin{equation}
\tilde\kappa_2=1.
\end{equation}
Since $\kappa_2 = \mu_2$  it is clear that eq. \eref{eqn:CumToMomConvert} is also
valid for the rescaled moments and cumulants. Because of the  
$N$ dependence eq. \eref{eqn:scaledCumuOrder}, the corrections of order
$1/N^{k-1}$ to {\it all} scaled moments 
only receive contributions from cumulants up to $\tilde \kappa_{2k}$, which
by themselves are completely determined by the moments up to order
$2k$ due to eq. \eqref{eqn:MomToCumConvert}. Hence we conclude:
\begin{center}
\emph{To order $N^{-k+1}$, all $\tilde M_{2p}$ are determined by a finite number of moments up to $\tilde M_{2l}$ ($l\leq k$), expanded to $N^{-k+1}$}.
\end{center}

For example, the $1/N^2$ expansion of all  moments is completely determined
by $\tilde \kappa_4$ and $\tilde \kappa_6$ which in turn are determined by $\tilde{M_4}$ and $\tilde M_6$, while at order $1/N^3$
the moments receive
only contributions from $\tilde \kappa_4$, $\tilde \kappa_6$  and
$\tilde \kappa_8$ and are thus determined by $\tilde M_4$, $\tilde M_6$ and $\tilde M_8$.
This is very surprising, because for large $p$ the Wick contractions contributing to the $2p$-th moment become rather complicated,
whereas to calculate up to $\tilde M_8$ we only need to consider a small set of
contraction diagrams. This will have important implications when it comes to the enumeration of intersection graphs, which is the subject of section \ref{sec:enumeration}.

The  scaled cumulant $\tilde \kappa_{2k}$ is $O(1/N^{k-1})$, but after being rescaled back to $\kappa_{2k}$ its leading term contributes
to the thermodynamic limit of the free energy. The leading term of $\tilde \kappa_{2k}$ is  determined
by the  $1/N^{k-1}$ corrections of the moments  up to order $2k$. We thus emphasize:
\begin{center}
  {\it The leading term of the $2k$-th scaled cumulant $\tilde \kappa_{2k}$ is determined by the moments $\tilde M_{2l}$ ($l\leq k$), expanded to order $N^{-k+1}$. The result for $\tilde \kappa_{2k}$ is valid
  for arbitrary $q$.}
\end{center}
This in particular implies that even if we want the complete information of
only the thermodynamic limit (leading in $1/N$) of the free energy, we would still need
all-order knowledge of scaled cumulants and hence of the scaled moments.

We will see below that the full $q$-dependence of the leading term of the sixth cumulant follows from
the $1/N^2$ corrections to the fourth and  sixth moment. The correction factor
is simply given by  $1-1/3q$ with no other corrections from higher moments.
So the large $q$ expansion of this cumulant terminates at this order.

The discussion in this section is general and also applies to the odd $q$ case, where only some minor changes of notations are needed.
\subsubsection{Explicit results to $1/N^3$}
\label{sec:3.2.2}
We will derive the expansion of  $\tilde M_{2p}$ to $1/N^3$ in this paper.
To the relevant order, eq. \eref{eqn:CumToMomConvert} can
be explicitly written as 
\begin{equation}\label{eqn:m2pInCumu}
\frac{\tilde M_{2p}}{(2p-1)!!} =1 + \frac{1}{3}\binom{p}{2}\tilde \kappa_4+\frac{1}{15}\binom{p}{3}\tilde \kappa_6+\frac{1}{3}\binom{p}{4} \tilde \kappa_4^2+\frac{1}{105}\binom{p}{4} \tilde \kappa_8 + \frac{2}{9}\binom{p}{5}\tilde\kappa_6\tilde\kappa_4+\frac{5}{9}\binom{p}{6}\tilde\kappa_4^3+O(1/N^4).
\end{equation}
We emphasize again the fact that only a finite number of terms need to be considered on the right-hand side of the above equation is due to the extensivity of
the free energy, see  eqs. \eqref{eqn:CumAsym} and \eqref{eqn:scaledCumuOrder}. 
The first eight moments were calculated exactly in \cite{GG-J-V}, and their
expansion up to order $1/N^3$ thus gives the expansion of all moments to this
order.

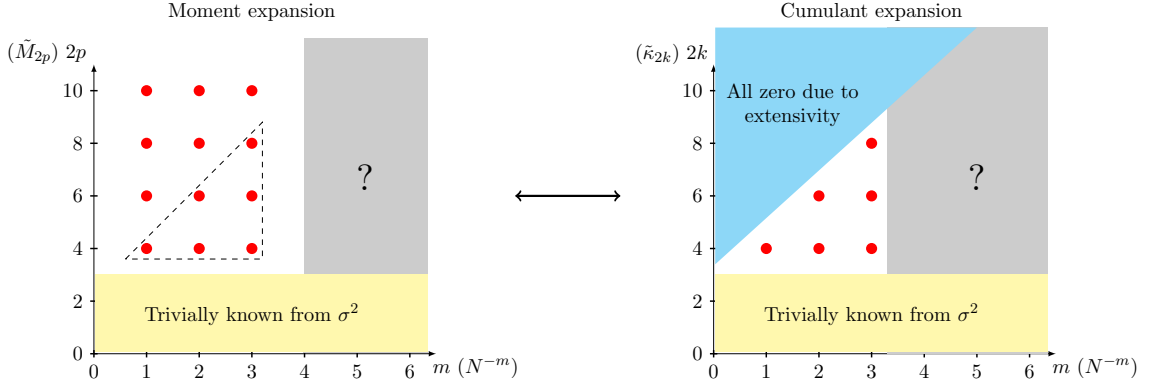
\begin{figure}
\begin{center}
\scalebox{.7}{
\begin{tikzpicture}
\node at (3, 6.5) {Moment expansion};
  \tkzInit[xmax=6,ymax=10,xmin=0,ymin=0, ystep=2]
   \tkzAxeXY[label={}]
  \tkzText[above](-0.9,10.75){$(\tilde M_{2p})\ 2p$}
    \tkzText[](7.25,-0.5){$m\ (N^{-m})$}
        \fill[gray!40] (4,0.03) rectangle (6.35,6);
                  \fill[yellow!40] (0.03,0.03) rectangle (6.35,1.5); 
                  \node at (3, 0.75) {Trivially known from $\sigma^2$};
   
        \draw[dashed] (0.6, 1.8)--(3.2, 4.4) -- (3.2, 1.8)--(0.6, 1.8);
     \fill[color=red] (1,2) circle(3pt);
       \fill[color=red] (1,3) circle(3pt);
     \fill[color=red] (1,4) circle(3pt);
          \fill[color=red] (1,5) circle(3pt);
        
               \fill[color=red]  (2,2) circle(3pt);
                 \fill[color=red]  (2,3) circle(3pt);  \fill[color=red]  (2,4) circle(3pt);  
                 \fill[color=red]  (2,5) circle(3pt);
    
      \fill[color=red] (3,2) circle(3pt);
        \fill[color=red] (3,3) circle(3pt);
         \fill[color=red] (3,4) circle(3pt);
         \fill[color=red] (3,5) circle(3pt);
       
          \node at (5.15,3.35) {\scalebox{1.8}{?}};
          \draw[<->, very thick]  (8,3)--(10,3);

\end{tikzpicture}          
          
\begin{tikzpicture}
\node at (3, 6.5) {Cumulant expansion};
           \tkzInit[xmax=6,ymax=10,xmin=0,ymin=0, ystep=2]
   \tkzAxeXY[label={}]
  \tkzText[above](-0.8,10.75){$(\tilde \kappa_{2k})\ 2k$}
    \tkzText[](7.25,-0.5){$m\ (N^{-m})$}

        \fill[color=red] (1,2) circle(3pt); 
        \fill[color=red] (2,2) circle(3pt); 
       \fill[color=red] (2,3) circle(3pt); 
       \fill[color=red] (3,2) circle(3pt); 
              \fill[color=red] (3,3) circle(3pt); 
                     \fill[color=red] (3,4) circle(3pt); 
            \fill[gray!40]  (3.3,0) -- (6.35,0) -- (6.35,6.2)--(3.3,6.2) -- cycle;
             \fill[cyan!40]  (0.03,1.7) -- (5,6.2) --(0.03,6.2)-- cycle;
      \fill[yellow!40] (0.03,0.03) rectangle (6.35,1.5); 
                  \node at (3, 0.75) {Trivially known from $\sigma^2$};
                  \node at (1.5,5){All zero due to};
                  \node at (1.5,4.5){ extensivity};
                  
          \node at (5,3.35) {\scalebox{1.8}{?}};
\end{tikzpicture}
}
\caption{Relation between two expansions (for even $q$). The horizontal axes denote orders in $1/N$, while the vertical axes denote orders in (scaled) moments and cumulants. The red dots are where the corresponding $1/N$ coefficients are known, and the gray region is unknown. It is interesting to see that on the moment side infinitely many coefficients are known, but they are all determined by the finite number of dots in the dashed triangle. This is ultimately because on the cumulant side we easily know an infinite number of coefficients due to extensivity (the cyan region).}\label{fig:knowledgeMap}
\end{center}
\end{figure}

Expanding the binomials in 
the results of \cite{GG-J-V} in powers of $1/N$ we obtain
\be
\begin{split}
\tilde M_4   =& 3-\frac{2 q^2}{N}+\frac{2 q^2(q-1)^2 }{N^2}-\frac{2 q^2 (q-1)^2 (2 q^2-8 q+5)}{3 N^3}+O\left(\frac{1}{N^4}\right),\\
 \tilde M_6=& 
 15-\frac{30q^2}{N}+2 q^2 \left(27 q^2-34 q+15\right)\frac{1}{N^2} \\
 &-2  q^2 (q-1)^2\left(38 q^2-56 q+25\right)\frac{1}{N^3}+O(N^{-4}),\\
 \tilde M_8=&
 105-\frac{420q^2}{N}+28 q^2 \left(44 q^2-38 q+15\right)\frac{1}{N^2} \\
&-4 q^2 \left(716 q^4-1712 q^3+1743 q^2-854 q+175\right)\frac{1}{N^3}+O(N^{-4}).
\end{split}
\ee 
This results in the cumulants:
\begin{equation}\label{eqn:cumuNcubeEvenq}
\begin{split}
\tilde \kappa_4=&-3+\tilde M_4\\
=&-\frac{2 q^2}{N}+\frac{2 q^2(q-1)^2 }{N^2}-\frac{2 q^2 (q-1)^2 (2 q^2-8 q+5)}{3 N^3}+O\left(N^{-4}\right),\\
\tilde \kappa_6=&30-15\tilde M_4+\tilde M_6\\
=&\frac{8 q^3 (3 q-1)}{N^2}-\frac{8 q^3(q-1)^2  (7 q-4)}{N^3}+O\left(N^{-4}\right),\\
\tilde \kappa_8=&-630+420\tilde M_4-35\tilde M_4^2-28\tilde M_6+\tilde M_8\\
=&-\frac{16 q^4 \left(46 q^2-36 q+7\right)}{N^3}+O\left(N^{-4}\right),
\end{split}
\end{equation}
which have exactly the leading order $N$ dependence of eq. \eqref{eqn:scaledCumuOrder} required for an
extensive free energy.
As we discussed at the begining of this section, these cumulants determine {\it all} moments to
order $1/N^3$.
Substituting these results for cumulants back to eq. \eref{eqn:m2pInCumu}, we thus obtain the following large $N$ expansion for the scaled moments to $1/N^3$:
\begin{equation}\label{eqn:m2pNcubeEvenq}
\begin{split}
\frac{\tilde M_{2p}}{(2p-1)!!} =&   1-\frac{2}{3} \binom{p}{2} \frac {q^2}{N}
  +\left[\binom{p}{2}\left(\frac{2}{3}q^2(q-1)^2\right)+\binom{p}{3}\left(\frac{8}{15}q^3(3q-1)\right)+\binom{p}{4}\frac{4}{3}q^4\right]\frac{1}{N^2}\\
  &-\left[\binom{p}{2}\left(\frac{2}{9}q^2(q-1)^2(2q^2-8q+5)\right)
+\binom{p}{3}\left(\frac{8}{15}q^3(q-1)^2(7q-4)\right)\right.\\
&\quad\ \ +\binom{p}{4} \left(\frac{8}{105}q^4(127q^2-142q+49)\right)+ \binom{p}{5}\left(\frac{32}{9}q^5(3q-1)\right)\\
&\quad\ \ \left.+\binom{p}{6}\frac{40}{9}q^6\right]\frac{1}{N^3}+O\left(\frac{1}{N^4}\right).
  \end{split}
\end{equation}
In \cite{GG-J-V} $\tilde M_{2p}$ was expanded to $1/N^2$ by Q-Hermite approximation and  triangle counting, and it agrees with \eref{eqn:m2pNcubeEvenq}, which we just obtained by a completely independent method.
The discussion above suggests an interesting ``map of knowledge'' between the moment expansion and cumulant expansion, as shown in figure \ref{fig:knowledgeMap}.

\subsubsection{Free energy}\label{sec:evenqFreeE}

We can substitute eq. \eqref{eqn:cumuNcubeEvenq} into eq. \eqref{eqn:HighTempCum} 
to obtain the high temperature expansion of the free energy: 
\begin{equation}
-\beta F= \frac{N}{2} \log 2 +\frac{1}{2!}\sigma^2 \beta^2+ \frac{1}{4!}\tilde \kappa_4 \sigma^4\beta^4+ \frac{1}{6!}\tilde \kappa_6 \sigma^6\beta^6 + \frac{1}{8!}\tilde \kappa_8 \sigma^8\beta^8+\cdots
\end{equation}
Since $\sigma^2 \sim N$  we obtain in the thermodynamic limit
\begin{equation}\label{eqn:freeEnergyHighTempNcubeEvenq}
\begin{split}
\frac{-\beta F}{N}= &\frac 12 \log 2 +\frac 1{4 q^2}  \beta ^2 -\frac{1}{2 q^2 4!}  \beta^4
+ \frac{8q^3(3q-1)}{8 q^6 6!}\beta^6 - \frac{16q^4(46q^2-36q+7)}{16 q^8 8!} \beta^8 +O(\beta^{10}),\\
\end{split}
\end{equation}
where we have set $J^2 = 2^{q-1}/q$. We stress that the $q$-dependence of the coefficient of
$\beta^{2k}$ is exact. In particular, it is valid for $q=2$ which can be easily shown by inserting
the result for the $q=2$ propagator (eq. (2.24) of \cite{maldacena2016})
\be
G_\beta(\tau) =\int_0^\pi\frac{d\theta}{\pi} \cos^2\theta
\frac{\cosh((\frac\tau\beta-\frac 12)2J\beta\sin\theta)}{\cosh (J\beta\sin\theta)}
  \ee
  into the expression for the free energy (eq. (2.27) of \cite{maldacena2016})
  \be
  J\partial_J (-\beta F/N)= -\frac \beta q \partial_\tau G|_{\tau\to 0}.
  \ee
Using the $1/N$ corrections
to the cumulants and the variance, it is straightforward to calculate $1/N$
corrections to the free energy. We have sufficient data to obtain terms up to $1/N^3$, each expanded to $\beta^8$, but since they are of limited physical
relevance, we have not written them down (for more details see appendix \ref{appen:freeEnergyNcube}).
To summarize, we have obtained the high temperature expansion of the
free energy to order $\beta^8$. 

Using a completely different method, a recent publication \cite{Tarnopolsky}
computes the large $q$ expansion to order $1/q^3$ for the free energy at leading order of $1/N$.\footnote{Their result, however, is valid for all temperatures.}  In addition to reproducing the large $q$ expansion calculated by \cite{Tarnopolsky} to order $1/q^3$ (see appendix \ref{append:tarnopolsky}), we also
obtain the $1/q^4$ correction to the free energy at order $\beta^8$ which is given by
\be
-N \frac 7{8! q^4} \beta^8.
\ee
Our results also show that there are no further large $q$ corrections for terms up to order $\beta^8$.

Since $\beta$ only appears in the combination $\beta J$, the high temperature expansion is the
weak coupling expansion in terms of Feynman graphs, where the full propagator is expanded in powers of $J$ with the free
propagator as the bare propagator. This way one can easily obtain the $\beta^2$ and $\beta^4$ corrections
to the free energy, but higher orders become much more tedious.

 We also remark that, if all one wants is the high temperature expansion of the free energy to a certain order in $\beta$ and $1/N$, a general expression for the large $N$ expansion of $\tilde M_{2p}$ such as eq. \eqref{eqn:m2pNcubeEvenq} is not necessary, as only equations such as \eqref{eqn:cumuNcubeEvenq} are used to calculate the high temperature expansion, which means only a finite number of moments are needed. However, the scope of this paper is wider than just computing the high temperature expansion, and a general expression for all $\tilde M_{2p}$ will be needed to solve the enumeration problem for intersection graphs, which will be discussed in section \ref{sec:enumeration}.

\subsection{Odd $q$ case}
We can repeat the same calculation for supersymmetric SYK models. However in this case
the Hamiltonian is given by the square of the supercharge so that the odd moments do not vanish.
Therefore, we need to define the scaled cumulants as 
\begin{equation}
\tilde \kappa_p :=\frac{\kappa_p}{\kappa_1^p}=\frac{\kappa_p}{M_2^p},
\end{equation}
because $M_{2}=\mu_1=\kappa_1$ for the odd $q$ case. This means 
\begin{equation}\label{eqn:scaledCumuOrderOddq}
\tilde \kappa_{k}\sim O(N^{-k+1}).
\end{equation}
As was the case for even $q$, a small number of low-order cumulants determine all moments to a
given order in $1/N$.
Repeating the same calculation that led to eq. \eqref{eqn:m2pInCumu}  we get 
\begin{equation}\label{eqn:m2pInCumuOddq}
\tilde M_{2p}=\frac{\mu_p}{\mu_1^p}=1+\binom p 2 \tilde\kappa_2+ \binom p 3 \tilde \kappa_3 + 3\binom p 4 \tilde \kappa_2^2 + \binom p 4 \tilde\kappa_4 +10\binom p 5 \tilde \kappa_2\tilde\kappa_3 + 15 \binom p 6 \tilde \kappa_2^3+O(N^{-4}).
\end{equation}
There is no $(2p-1)!!$ factor this time. To calculate $\tilde \kappa_2$, $\tilde \kappa_3$ and $\tilde \kappa_4$ we will need $\tilde M_4$, $\tilde M_6$
 and $\tilde M_8$ as we did for the even $q$ case. They can be again calculated from the analytical formulas in \cite{GG-J-V} resulting in
 \begin{equation}
 \begin{split}
\tilde M_4  =& 1+\frac{2 q^2}{N} -2q^2 (q-1)^2 \frac{1}{N^2}+\frac{2}{3}q^2  (q-1)^2 \left(2 q^2-8 q+5\right) \frac{1}{N^3}+O\left(N^{-4}\right),\\
\tilde M_6=&1+\frac{6q^2}{N} -2 q^2 (q-3) (3 q-1)\frac{1}{N^2}+2 q^2 (q-1)^2 \left(6 q^2-24 q+5\right)\frac{1}{N^3}+O\left(N^{-4}\right),\\
\tilde M_8=&1+\frac{12q^2}{N}+4 q^2 (14 q-3)\frac{1}{N^2}-4 q^2 \left(4 q^4+32 q^3-93 q^2+50 q-5\right)\frac{1}{N^3}+O\left(N^{-4}\right).
\end{split}
\end{equation}
Using the relation between cumulants and moments, eq. \eqref{eqn:MomToCumConvert}, we obtain 
\begin{equation}\label{eqn:cumulargeNoddq}
\begin{split}
\tilde \kappa_2 &= \tilde \mu_2 -1 = \tilde M_4-1 \\
&=\frac{2 q^2}{N} -2q^2 (q-1)^2 \frac{1}{N^2}+\frac{2}{3} q^2(q-1)^2  \left(2 q^2-8 q+5\right) \frac{1}{N^3}+O\left(N^{-4}\right),\\
\tilde \kappa_3 &= \tilde M_6 -3 \tilde M_4 +2 \\
&= \frac{8q^3}{N^2}+8 q^3(q-4) (q-1)^2 \frac{1}{N^3}+O\left(N^{-4}\right),\\
\tilde \kappa_4 &= \tilde M_8 -4\tilde M_6-3\tilde M_4^2+12\tilde M_4-6 \\
&=-16 q^4 \left(2 q^2-4 q-1\right)\frac{1}{N^3}+O\left(N^{-4}\right).
\end{split}
\end{equation}
Together with \eqref{eqn:m2pInCumuOddq}, we conclude that the $p$-dependence of the moments is given by
\begin{equation}
\begin{split}
\tilde M_{2p}=&1+\binom p 2 \frac{2 q^2}{N}-\left[\binom p 2 2q^2(q-1)^2-\binom p 3 8q^3-\binom p 4 12q^4\right]\frac{1}{N^2}\\
&+ \left[\binom p 2\frac{2}{3} q^2(q-1)^2  \left(2 q^2-8 q+5\right)+\binom p 3 8 q^3 (q-1)^2(q-4)\right.\\
&\quad\ \  \left.-\binom p 4 8 q^4 \left(7 q^2-14 q+1\right)+\binom p 5 160q^5+\binom p 6 120q^6 \right]\frac{1}{N^3}
+O(N^{-4}).
\end{split}
\end{equation}
The high temperature expansion of free energy has the form:
\begin{equation}
-\beta F= \frac{N}{2} \log 2 -\sigma^2 \beta+ \frac{1}{2!}\sigma^4 \tilde\kappa_2\beta^2 -\frac{1}{3!}\sigma^6 \tilde\kappa_3\beta^3+\frac{1}{4!}\sigma^8 \tilde\kappa_4\beta^4+O(\beta^5).
\end{equation}
More explicitly, we have in the thermodynamic limit
\begin{equation}
\begin{split}\label{eqn:freeEnergyHighTempNcube}
\frac{-\beta F}{N}= &\frac 12 \log 2-\frac 1{2 q^2}  \beta  +\frac{1}{2!(2 q^2) }  \beta^2
- \frac{1}{3! q^3 }\beta^3 + \frac{-(2q^2-4q-1)}{ 4!q^4} \beta^4 +O(\beta^{5}),
\end{split}
\end{equation}
where again we have set $J^2 = 2^{q-1}/q$. We only displayed the coefficient of the leading term in $1/N$, but the $1/N$ corrections can be calculated in a straightforward way as well (see appendix \ref{appen:freeEnergyNcube}).

\section{Enumerative identities of intersection graphs}\label{sec:enumeration}

In this section we derive enumerative identities for intersection graphs, some of which have
not appeared in the literature.
We start with a short review of the graphical calculation of
the moments, and the details can be found in \cite{GG-J-V}.

\subsection{Graphical calculation of moments}

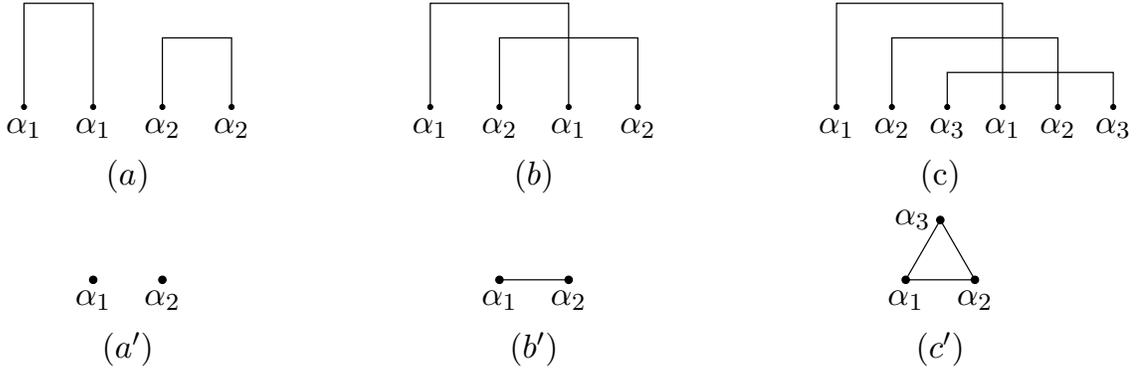
\begin{figure}[t!]
  \begin{center}
\begin{equation}
    \resizebox{\textwidth}{!}{
  \begin{tikzpicture}[scale=0.8]
\draw[fill=black] (0,0) circle (1pt);
\draw[fill=black] (1,0) circle (1pt);
\draw[fill=black] (2,0) circle (1pt);
\draw[fill=black] (3,0) circle (1pt);

\node at (0,-0.3) {$\alpha_1$};
\node at (1,-0.3) {$\alpha_1$};
\node at (2,-0.3) {$\alpha_2$};
\node at (3,-0.3) {$\alpha_2$};

\node at (1.5,-1) {($a$)};

\draw[fill=black] (1,-2.5) circle (1.5pt);
\draw[fill=black] (2,-2.5) circle (1.5pt);
\node at (1, -2.8 )  {$\alpha_1$};
\node at (2, -2.8 )  {$\alpha_2$};
\node at (1.5,-3.5) {($a'$)};
\draw[] (0,0) -- (0,1.5) -- (1,1.5) -- (1,0);
\draw[] (2,0) -- (2,1) -- (3,1) -- (3,0);
\end{tikzpicture}\nn 
\hspace*{1.5cm}
\begin{tikzpicture}[scale=0.8]
\draw[fill=black] (0,0) circle (1pt);
\draw[fill=black] (1,0) circle (1pt);
\draw[fill=black] (2,0) circle (1pt);
\draw[fill=black] (3,0) circle (1pt);

\node at (0,-0.3) {$\alpha_1$};
\node at (1,-0.3) {$\alpha_2$};
\node at (2,-0.3) {$\alpha_1$};
\node at (3,-0.3) {$\alpha_2$};

\node at (1.5,-1) {($b$)};

\draw[fill=black] (1,-2.5) circle (1.5pt);
\draw[fill=black] (2,-2.5) circle (1.5pt);
\node at (1, -2.8 )  {$\alpha_1$};
\node at (2, -2.8 )  {$\alpha_2$};
\draw[] (1,-2.5)--(2,-2.5);
\node at (1.5,-3.5) {($b'$)};
\draw[] (0,0) -- (0,1.5) -- (2,1.5) -- (2,0);
\draw[] (1,0) -- (1,1) -- (3,1) -- (3,0); 

\end{tikzpicture}\nn 
\hspace*{1.5cm}
\begin{tikzpicture}[scale=0.8]
\draw[fill=black] (0,0) circle (1pt);
\draw[fill=black] (0.8,0) circle (1pt);
\draw[fill=black] (1.6,0) circle (1pt);
\draw[fill=black] (2.4,0) circle (1pt);
\draw[fill=black] (3.2,0) circle (1pt);
\draw[fill=black] (4,0) circle (1pt);

\node at (0,-0.3) {$\alpha_1$};
\node at (0.8,-0.3) {$\alpha_2$};
\node at (1.6,-0.3) {$\alpha_3$};
\node at (2.4,-0.3) {$\alpha_1$};
\node at (3.2,-0.3) {$\alpha_2$};
\node at (4,-0.3) {$\alpha_3$};

\node at (1.5,-1) {(c)};
\draw[fill=black] (1,-2.5) circle (1.5pt);
\draw[fill=black] (2,-2.5) circle (1.5pt);
\draw[fill=black] (1.5,-1.634) circle (1.5pt);

\node at (1, -2.8 )  {$\alpha_1$};
\node at (2, -2.8 )  {$\alpha_2$};
\node at (1.1, -1.634 )  {$\alpha_3$};

\node at (1.5,-3.5) {($c'$)};
\draw[] (0,0) -- (0,1.5) -- (2.4,1.5) -- (2.4,0);
\draw[] (0.8,0) -- (0.8,1) -- (3.2,1) -- (3.2,0); 
\draw[] (1.6,0) -- (1.6,0.5) -- (4,0.5) -- (4,0); 
\draw[] (1,-2.5)-- (2,-2.5) -- (1.5,-1.634) -- (1,-2.5); 

\end{tikzpicture}
}
\end{equation}
\end{center}
\caption{Three examples of  chord diagrams ($a$),($b$), ($c$), and the corresponding intersection graphs ($a'$), ($b'$) and ($c'$). In particular ($b$) and ($b'$) represent the value of $\eta$ in eq. \eqref{eqn:etaDef}. } \label{fig:m4Diagrams}
\end{figure}

The moments of the SYK model are given by the expectation value
\be
\left \langle 2^{-N/2}\Tr( J_\alpha \Gamma_\alpha)^{2p}\right \rangle.
\ee
Because the probability is Gaussian, the average is given by the sum
over all possible Wick contractions, which can be represented
by \textit{rooted chord diagrams}, and some examples of such chord diagrams
are given in the first row of figure \ref{fig:m4Diagrams}. For large $N$ and finite $p$, the indices of the
$\Gamma_\alpha$ are mostly different, so that they can be commuted to
pairs of $\Gamma_\alpha\Gamma_\alpha =1$ (no implicit summation over $\alpha$). Since in this limit all contractions contribute
equally for even $q$, this results in a Gaussian spectral density. For odd
$q$, when two $\Gamma_\alpha$ and $\Gamma_\beta$ with no common indices
anti-commute, the contractions are alternately positive and negative, leaving
only one net contraction for all moments which results in the moments of
two delta functions in this limit. We always consider the scaled moments $\tilde M_{2p}:=M_{2p}/M_2^p$, so that
the variance cancels in the ratio, and the values of chord diagrams always refer to the values of Wick contractions normalized by $M_2^p$.

To calculate $1/N$ corrections, we have to take into account that
$\Gamma_\alpha $ and $\Gamma_\beta$ commute or anti-commute depending
on how many indices they have in common,
\be
\Gamma_\alpha \Gamma_\beta+(-1)^{q+1+k}\Gamma_\beta \Gamma_\alpha=0,
\ee
where $k$ is the number of common indices in $\alpha$ and $\beta$.
Taking this into account, we obtain the value of two intersecting contraction
lines:
\be\label{eqn:etaDef}
\eta=(-1)^q {N \choose q}^{-1} \sum_{k=0}^q (-1)^k{N-q \choose q-k} {q\choose k}.
\ee
We can further translate chord diagrams into \textit{intersection graphs}, which are obtained by representing each chord by a vertex, and connecting two vertices if and only if the two chords they represent intersect each other in the chord diagram. We give some examples of intersection graphs in the second row of figure \ref{fig:m4Diagrams}. We will denote a generic intersection graph by $G$.
An important approximation to the moments is to treat all intersections of
contraction lines as independent. In the language of intersection graphs, if an intersection graph $G$ has
$E$ edges, its value $\eta_G$ is approximated by
\be\label{eqn:QHforEta}
\eta_G \approx \eta^E.
\ee
This  approximation  gives  moments that are correct up to order $1/N$.
The corresponding moments are the moments of the
weight function of the Q-Hermite polynomials. That is why this approximation
is known as the Q-Hermite approximation. The Q-Hermite approximation to moments is thus the sum of $\eta^E$ over all the $(2p-1)!!$ intersection graphs.

For more details of the graphical calculations including the graph-theoretic
identities needed to sum all graphs we refer to \cite{GG-J-V}. To conclude this review, we remark that the method of \cite{GG-J-V} relies on a back-and-forth interplay between the moment expansion and intersection graphs: intersection graphs inform us on how to calculate moments, and moment calculations for the
exactly solvable $q=1$ and $q=2$ SYK model prove enumerative identities about intersection graphs, which in turn feed back to the moment calculation for general $q$. In the current paper the relation is more one-way: we have obtained results for general moments in previous sections without relying on the enumeration of intersection graphs, and the graphs at best could play a minor role as a book keeping device. In any case, now we can use the results for the moment expansion to prove identities for intersection graphs, which will be discussed in this section.  

\subsection{Structure of contributions}

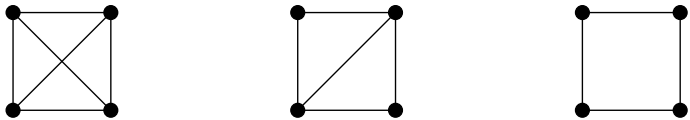
\begin{figure}
\begin{center}
\scalebox{1.3}{
\begin{tikzpicture}
\draw[fill=black] (0,0) circle (2pt);
\draw[fill=black] (1,0) circle (2pt);
\draw[fill=black] (1,1) circle (2pt);
\draw[fill=black] (0,1) circle (2pt);
\draw[](0,0) -- (0,1)--(1,1)--(1,0)--(0,0)--(1,1);
\draw[](1,0) -- (0,1);
\end{tikzpicture}
\hspace{1.5cm}
\begin{tikzpicture}
\draw[fill=black] (2,0) circle (2pt);
\draw[fill=black] (3,0) circle (2pt);
\draw[fill=black] (3,1) circle (2pt);
\draw[fill=black] (2,1) circle (2pt);

\draw[](2,0) -- (2,1)--(3,1)--(3,0)--(2,0)--(3,1);
\end{tikzpicture}
\hspace{1.5cm}
\begin{tikzpicture}
\draw[fill=black] (4,0) circle (2pt);
\draw[fill=black] (5,0) circle (2pt);
\draw[fill=black] (5,1) circle (2pt);
\draw[fill=black] (4,1) circle (2pt);

\draw[](4,0) -- (4,1)--(5,1)--(5,0)--(4,0);
\end{tikzpicture} 
}
\end{center}
\caption{The four-point structures appearing in a general intersection graph $G$. Their numbers are denoted by $f_6$, $f_5$ and $f_4$  from the left to the right, respectively.}\label{fig:4pointStructures}
\end{figure}

In section 3, we have calculated the large $N$ expansions of $\tilde M_{2p}$ to order $1/N^3$.
We did not use any enumerative identities like the ones in \cite{GG-J-V}. Nevertheless, we see that various
binomial factors arise in eqs. \eqref{eqn:m2pInCumu} and \eqref{eqn:m2pInCumuOddq} simply from the relations between the moments and the cumulants.
This suggests we can turn around and use the calculations we just presented
to generate  enumerative identities for intersection graphs.
To see what type of graph-theoretic objects are to be enumerated,
we first state the main
result of the next two sections:
\be
\label{eqn:etaDiffNcube}
\eta_G -\eta^E&=&(-1)^{Eq}\frac{-8q^3}{N^2} T
 \\ &&
+ (-1)^{Eq}[ 16ETq^5+(-72T-80f_6-16f_5+16f_4)q^4+32Tq^3]\frac 1 {N^3}+ O(N^{-4}),\nn
\ee
where $E$ is the numbers of edges, $T$ is the number of triangles, and $f_6$, $f_5$, $f_4$ are the number of  the four-point structures
depicted in figure \ref{fig:4pointStructures}, in an intersection graph $G$.
We can check this formula for a few  nontrivial graphs that contribute to moments up to $\tilde M_8$ denoted by $\eta,\; T_6,\; T_{44},\; T_{66}$ and $T_8$ in tables \ref{tab:6thMomIntersecGraphs} and \ref{tab:8thMomIntersecGraphs}. Expanding
 the results for these graphs obtained in \cite{GG-J-V} to order $1/N^3$ we obtain,
  \begin{equation}\label{exps}
   \begin{split}
     (-1)^q\eta&=1-\frac{2 q^2}{N}+\frac{2 q^2(q-1)^2 }{N^2}-\frac{2 q^2 (q-1)^2 (2 q^2-8 q+5)}{3 N^3}+O\left(\frac{1}{N^4}\right)\\
(-1)^{3q}T_6&= (-1)^{3q}\eta^3-\frac{8q^3}{N^2}+(48q^5-72q^4+32q^3)\frac{1}{N^3}+O\left(\frac{1}{N^4}\right),\\
(-1)^{4q}T_{44}&=(-1)^{4q}\eta^4+\frac{16q^4}{N^3}+O\left(\frac{1}{N^4}\right),\\
(-1)^{5q}T_{66}&=(-1)^{5q}\eta^5-\frac{16q^3}{N^2}+(160q^5-160q^4+64q^3)\frac{1}{N^3}+O\left(\frac{1}{N^4}\right),\\
(-1)^{6q}T_{8}&=(-1)^{6q}\eta^6-\frac{32q^3}{N^2}+(384q^5-368q^4+128q^3)\frac{1}{N^3}+O\left(\frac{1}{N^4}\right).\\
\end{split}
\end{equation}
\begin{table}[t!]
  \begin{center}
\begin{tabular}{|c|c|c|c|c|}
\hline Intersection graph & \begin{tikzpicture} \draw[fill=black] (0,0) circle (1pt); \draw[fill=black] (0.4,0) circle (1pt);\draw[fill=black] (0.2,0.346) circle (1pt); \node at (0.2,0.35) {};\end{tikzpicture} & \begin{tikzpicture} \draw[fill=black] (0,0) circle (1pt); \draw[fill=black] (0.4,0) circle (1pt);\draw[fill=black] (0.2,0.346) circle (1pt);
\draw (0,0)--(0.4,0); \end{tikzpicture} & \begin{tikzpicture} \draw[fill=black] (0,0) circle (1pt); \draw[fill=black] (0.4,0) circle (1pt);\draw[fill=black] (0.2,0.346) circle (1pt);
\draw (0,0)--(0.4,0)--(0.2,0.346); \end{tikzpicture} & \begin{tikzpicture} \draw[fill=black] (0,0) circle (1pt); \draw[fill=black] (0.4,0) circle (1pt);\draw[fill=black] (0.2,0.346) circle (1pt);
\draw (0,0)--(0.4,0)--(0.2,0.346)--(0,0); \end{tikzpicture} \\ \hline
Value & 1 & $\eta$ & $\eta^2$ & $T_6$\\ \hline
Multiplicity & 5 & 6 & 3 & 1\\ \hline
\end{tabular}
\caption{All the intersection graphs contributing to the sixth moment.}\label{tab:6thMomIntersecGraphs}
\end{center}
\end{table}

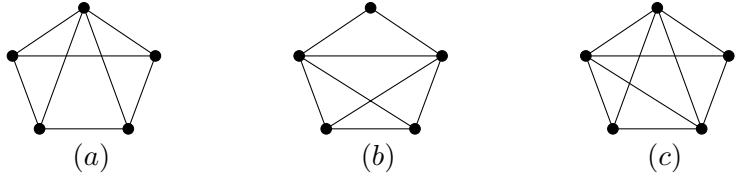
\begin{figure}
\begin{center}
\begin{tikzpicture}
\draw[fill=black] (0,0.8) circle (2pt);
\draw[fill=black] (0.95,0.16) circle (2pt);
\draw[fill=black] (0.59,-0.81) circle (2pt);
\draw[fill=black] (-0.59,-0.81) circle (2pt);
\draw[fill=black] (-0.95,0.16) circle (2pt);
\node at (0.1,-1.2) {$(a)$};

\draw (0,0.8) -- (0.95,0.16) -- (0.59,-0.81) -- (-0.59,-0.81) -- (-0.95,0.16) -- (0,0.8) -- (-0.59,-0.81);
\draw (0,0.8) -- (0.59,-0.81);
\draw (0.95,0.16)-- (-0.95,0.16);
\end{tikzpicture} 
\hspace{1.5cm}
\begin{tikzpicture}
\draw[fill=black] (0,0.8) circle (2pt);
\draw[fill=black] (0.95,0.16) circle (2pt);
\draw[fill=black] (0.59,-0.81) circle (2pt);
\draw[fill=black] (-0.59,-0.81) circle (2pt);
\draw[fill=black] (-0.95,0.16) circle (2pt);
\node at (0.1,-1.2) {$(b)$};
\draw (0,0.8) -- (0.95,0.16) -- (0.59,-0.81) -- (-0.59,-0.81) -- (-0.95,0.16) -- (0,0.8);
\draw (-0.59,-0.81) -- (0.95,0.16) -- (-0.95,0.16) -- (0.59,-0.81);
\end{tikzpicture}
\hspace{1.5cm}
 \begin{tikzpicture}
\draw[fill=black] (0,0.8) circle (2pt);
\draw[fill=black] (0.95,0.16) circle (2pt);
\draw[fill=black] (0.59,-0.81) circle (2pt);
\draw[fill=black] (-0.59,-0.81) circle (2pt);
\draw[fill=black] (-0.95,0.16) circle (2pt);
\node at (0.1,-1.2) {$(c)$};

\draw (0,0.8) -- (0.95,0.16) -- (0.59,-0.81) -- (-0.59,-0.81) -- (-0.95,0.16) -- (0,0.8) -- (-0.59,-0.81);
\draw (0,0.8) -- (0.59,-0.81);
\draw (0.95,0.16)-- (-0.95,0.16)--(0.59,-0.81) ;
\end{tikzpicture} 
\end{center}
\caption{Some examples of four-point structure counting in intersection graphs with five vertices: $(a)$ has $E=8,\; T=4$ and $f_6=0,\; f_5=4,\; f_4=1$; $(b)$ has $E=8,\; T=5$ and $f_6=1,\; f_5=2,\; f_4=0$; $(c)$ has $E=9,\; T=7$ and $f_6=2,\; f_5=3,\; f_4=0$.} \label{fig:penta}
\end{figure}

\begin{table}
  \begin{center}
\begin{tabular}{|c|c|c|c|c|c|c|c|c|c|c|c|}
\hline Intersection graph & \begin{tikzpicture} \draw[fill=black] (0,0) circle (1pt); \draw[fill=black] (0.4,0) circle (1pt);\draw[fill=black] (0.4,0.4) circle (1pt);\draw[fill=black] (0,0.4) circle (1pt); \node at (0.2,0.44) {};\end{tikzpicture} & \begin{tikzpicture}\draw[fill=black] (0,0) circle (1pt); \draw[fill=black] (0.4,0) circle (1pt);\draw[fill=black] (0.4,0.4) circle (1pt);\draw[fill=black] (0,0.4) circle (1pt);
\draw (0,0)--(0.4,0); \end{tikzpicture} & \begin{tikzpicture} \draw[fill=black] (0,0) circle (1pt); \draw[fill=black] (0.4,0) circle (1pt);\draw[fill=black] (0.4,0.4) circle (1pt);\draw[fill=black] (0,0.4) circle (1pt);
\draw (0,0)--(0.4,0);\draw (0,0.4)--(0.4,0.4); \end{tikzpicture} & \begin{tikzpicture} \draw[fill=black] (0,0) circle (1pt); \draw[fill=black] (0.4,0) circle (1pt);\draw[fill=black] (0.4,0.4) circle (1pt);\draw[fill=black] (0,0.4) circle (1pt);
\draw (0,0)--(0.4,0)--(0.4,0.4); \end{tikzpicture} & \begin{tikzpicture} \draw[fill=black] (0,0) circle (1pt); \draw[fill=black] (0.4,0) circle (1pt);\draw[fill=black] (0.4,0.4) circle (1pt);\draw[fill=black] (0,0.4) circle (1pt);
\draw (0,0)--(0.4,0)--(0.4,0.4);\draw (0.4,0)--(0,0.4); \end{tikzpicture}& \begin{tikzpicture} \draw[fill=black] (0,0) circle (1pt); \draw[fill=black] (0.4,0) circle (1pt);\draw[fill=black] (0.4,0.4) circle (1pt);\draw[fill=black] (0,0.4) circle (1pt);
\draw (0,0)--(0.4,0)--(0.4,0.4)--(0,0.4); \end{tikzpicture}& \begin{tikzpicture} \draw[fill=black] (0,0) circle (1pt); \draw[fill=black] (0.4,0) circle (1pt);\draw[fill=black] (0.4,0.4) circle (1pt);\draw[fill=black] (0,0.4) circle (1pt);
\draw (0,0)--(0.4,0)--(0.4,0.4)--(0,0); \end{tikzpicture}& \begin{tikzpicture} \draw[fill=black] (0,0) circle (1pt); \draw[fill=black] (0.4,0) circle (1pt);\draw[fill=black] (0.4,0.4) circle (1pt);\draw[fill=black] (0,0.4) circle (1pt);
\draw (0,0)--(0.4,0)--(0.4,0.4)--(0,0)--(0,0.4); \end{tikzpicture}& \begin{tikzpicture} \draw[fill=black] (0,0) circle (1pt); \draw[fill=black] (0.4,0) circle (1pt);\draw[fill=black] (0.4,0.4) circle (1pt);\draw[fill=black] (0,0.4) circle (1pt);
\draw (0,0)--(0.4,0)--(0.4,0.4)--(0,0.4)--(0,0); \end{tikzpicture}& \begin{tikzpicture} \draw[fill=black] (0,0) circle (1pt); \draw[fill=black] (0.4,0) circle (1pt);\draw[fill=black] (0.4,0.4) circle (1pt);\draw[fill=black] (0,0.4) circle (1pt);
\draw (0,0)--(0.4,0)--(0.4,0.4)--(0,0.4)--(0,0)--(0.4,0.4); \end{tikzpicture}& \begin{tikzpicture} \draw[fill=black] (0,0) circle (1pt); \draw[fill=black] (0.4,0) circle (1pt);\draw[fill=black] (0.4,0.4) circle (1pt);\draw[fill=black] (0,0.4) circle (1pt);
\draw (0,0)--(0.4,0)--(0.4,0.4)--(0,0.4)--(0,0)--(0.4,0.4);\draw (0,0.4)--(0.4,0); \end{tikzpicture}\\ \hline
Value & 1 & $\eta$ & $\eta^2$ & $\eta^2$ & $\eta^3$ & $\eta^3$ &$T_6$ & $\eta T_6$ & $T_{44}$& $T_{66}$ & $T_8$\\ \hline
Multiplicity & 14 & 28 & 4 & 24 & 4 & 8 & 8& 8& 2 & 4 &1\\ \hline
\end{tabular}
\caption{All the intersection graphs contributing to the eighth moment.}\label{tab:8thMomIntersecGraphs}
\end{center}
\end{table}
Using the graphs in tables \ref{tab:6thMomIntersecGraphs}
 and \ref{tab:8thMomIntersecGraphs}
one  can easily verify  that the above results satisfy eq. \eqref{eqn:etaDiffNcube}. 
Note that triangles (whose value is $T_6$) made their first appearance as a complete
intersection graph for the sixth moment (table \ref{tab:6thMomIntersecGraphs}), and in the eighth moment, they become substructures of various graphs.
The same is true for the four-point structures whose values are $T_{44}$, $T_{66}$ and $T_8$: they first appear as complete
    graphs for the eighth moment (table \ref{tab:8thMomIntersecGraphs}), and will become substructures for higher moments. Some examples are given in figure \ref{fig:penta}.

How do we proceed to prove eq. \eqref{eqn:etaDiffNcube} in its full generality? 
A rigorous proof of eq. \eqref{eqn:etaDiffNcube} to  order $1/N^2$ was given in \cite{GG-J-V}, which is
essentially a very tedious brute-force calculation. The $1/N^3$ term can be obtained rigorously by the same method, but the calculation turns out to be extremely tedious and uninstructive. So instead, in the next two sections,
we will present a less rigorous but more instructive method to obtain the result
\eref{eqn:etaDiffNcube}. As a warm-up we will
start with the $1/N^2$ term in the next section, and then continue with the $1/N^3$ corrections.

\subsection{Graphical calculation of the $1/N^2$ term}

We first take note of a rather trivial fact that if an intersection graph $G$
is disconnected with two components $G_1$ and $G_2$, then $\eta_G=\eta_{G_1}\eta_{G_2}$. Moreover,
this factorization property also holds in a less trivial situation, where an
intersection graph can be made disconnected by cutting a vertex, which was proved
in \cite{GG-J-V}.\footnote{In a nutshell, the cut-vertex factorization property holds because the subscripts $\alpha$ (vertices in intersection graphs) for the $\Gamma_\alpha$ live in a ``isotropic'' space, so that if one first sums over all vertices on one side of the cut-vertex, the result no longer depends on the cut-vertex. A graph that cannot be made disconnected by cutting a single vertex is called ``2-connected'' in mathematics literature, here we just call it ``irreducible''.} To be concrete, let us first look at the $1/N^2$ order corrections. 
We have argued that the $1/N^2$ coefficient for $\tilde M_{2p}$ is completely
determined by $\tilde M_4$ and $\tilde M_6$, which  only depend on the values of edges and triangles 
($\eta$ and $T_6$) according to table \ref{tab:6thMomIntersecGraphs} and factorization. However, this fact appears rather mysterious if one thinks about the moments in
terms of the Wick contractions
\begin{equation}
\tilde M_{2p} = \sum_{i=1}^{(2p-1)!!} \eta_{G_i},
\end{equation}
 where $G_i$ denotes the $i$-th intersection graph and all the intersection graphs have $p$ vertices. 
An intersection graph can get quite complicated when $p$ is large, for example imagine a complete graph with $p$ vertices, which has $\binom{p}{2}$ edges. On top of that one then needs to sum over all these complicated graphs. How can one tame such a complex beast by only using $\tilde M_4$ and $\tilde M_6$? 
The only plausible way to reconcile these two seemingly conflicting view points is that
the $1/N^2$ coefficient of each $\eta_{G_i}$ ought to be a function 
of only the number of edges and triangles in that intersection graph, not any other property of the graph. So the fact that we are only looking at a fixed order in $1/N$  allows a huge simplification to happen.
We may summarize this plausible result as
\begin{equation}\label{eqn:etaTrianUnderter}
 \eta_G- \eta^E = \frac{1}{N^2} A(E,T) + O(1/N^3),
\end{equation}
where $A(E,T)$ is some function of $E$ and $T$.\footnote{Eq. \eqref{eqn:etaTrianUnderter} takes for granted that $\eta_G- \eta^E$ is $1/N$-exact. It will be clear that the $1/N$-exactness can be proved by the method we are unfolding, but since $1/N$-exactness of $\eta_G- \eta^E$ was extensively discussed in \cite{garcia2016,garcia2017,GG-J-V}, we choose not to prove it in this discussion. } Note that since $\eta^E$ is a function of $E$, the plausibility argument loses no generality by considering $\eta_G- \eta^E$ instead of $\eta_G$ on the left-hand side of eq. \eqref{eqn:etaTrianUnderter}, and this is motivated by the Q-Hermite approximation eq. \eqref{eqn:QHforEta}.
Let us pause here and summarize the reasons for the necessity for such simplicity:
\begin{itemize}
\item \textit{The cut-vertex factorization property of $\eta_G$, which together with table \ref{tab:6thMomIntersecGraphs} implies $\tilde M_4$ and $\tilde M_6$ depend only on the values of edges and  triangles ($\eta$ and $T_6$).}
\item \textit{The extensivity of the free energy/cumulants, which forces the $1/N^2$ coefficient of $\tilde M_{2p}$ to depend on only $\tilde M_4$ and $\tilde M_6$, for any value of $p$.}
\end{itemize}

Now the task is to fix the explicit form of $A(E,T)$. It
is sufficient to consider those special graphs $G$ with only
disconnected irreducible structures, so that there is a complete factorization. Since the relevant irreducible structures can only be edges or triangles,
let us take $G$ to be an intersection graph with $E$ edges and $T$ triangles where all the triangles are disconnected, and the edges other than the ones that make up triangles are also disconnected, which means there are $E-3T$ of them. This implies
 \begin{equation}
 \eta_G = T_6 ^{T}\eta^{E-3T},
 \end{equation}
 where $T_6$ is the value of the Wick contraction represented by a triangle. 	We now have 
 \begin{equation}\label{eqn:traingleFactori}
   \eta_G- \eta^E =   \eta^E\left(\left [\frac{T_6}{\eta^3} \right]^{T} -1\right ).
 \end{equation}
 We define
\be
\delta T_6: = \frac{T_6}{\eta^3} - 1.
 \ee
By the expansion of $T_6$ in eq. \eqref{exps}, we have $\delta T_6 \sim O(1/N^2)$ so that
 \be
   \eta_G- \eta^E = T \eta^E\delta T_6 +O(1/N^3).
\ee
Using that
 \be
 \eta &=&(-1)^q + O(1/N),\\
 \delta T_6& =&\frac{-8q^3}{N^2} +O(1/N^3),
 \ee
 we finally obtain
 \begin{equation}
  \eta_G- \eta^E  = (-1)^{Eq}\left(\frac{-8q^3}{N^2}\right)T+O(1/N^3).
 \end{equation}
Hence by considering special graphs we have fixed the form of $A(E,T)$ to be $A(E,T)=(-1)^{Eq}(-8q^3)T$, which must also be true for a general graph. This proves eq. \eqref{eqn:etaDiffNcube} to order $1/N^2$.

\subsection{Graphical calculation of the $1/N^3$ term}
We may continue with this strategy to order $1/N^3$. Since $\tilde M_{2p}$ to this order is completely determined by $\tilde M_4$, $\tilde M_6$ and $\tilde M_8$, by the same argument as in previous subsection, we conclude for any intersection graph $G$ we have 
 \begin{equation}
  \eta_G- \eta^E  = (-1)^{Eq}\left(\frac{-8q^3}{N^2}\right)T+ \frac 1 {N^3} B(E,T, f_6, f_5, f_4)+O(1/N^4),
 \end{equation}
because $E,\; T,\; f_6,\; f_5$ and $f_4$ count all the irreducible structures that appear for moments up to $\tilde M_8$. Again consider an intersection graph $G$ with $E$ edges, $T$ triangles, and the four-point structures counted by $f_6,\; f_5$ and $f_4$ (see figure \ref{fig:4pointStructures}), where all the irreducible structures are disconnected, so we have $T-4f_6-2f_5$ disconnected triangles and $E-3(T-4f_6-2f_5)-6f_6-5f_5-4f_4=E-3T+6f_6+f_5-4f_4$ disconnected edges. Thus
\begin{equation}
\begin{split}
  \eta_G=&\eta^{E-3T+6f_6+f_5-4f_4}{T_6}^{T-4f_6-2f_5} {T_8}^{f_6}
  {T_{66}}^{f_5}{T_{44}}^{f_4}
  \\
  =&\eta^E \left[\frac{T_6}{\eta^3}\right]^{T-4f_6-2f_5}\left[ \frac{T_8}{\eta^6}\right]^{f_6}
  \left[\frac{T_{66}}{\eta^5}\right]^{f_5}
  \left[\frac{T_{44}}{\eta^4}\right]^{f_4}
  ,
  \end{split}
\end{equation}
where $T_8,\; T_{66}$ and $T_{44}$ are the values of the intersection graphs counted by $f_6, \;f_5$ and $f_4$, see table \ref{tab:8thMomIntersecGraphs}.
We have that
\be
\begin{split}
\eta =& (-1)^q(1 + \delta \eta) + O(1/N^2),\\
\frac {T_6}{\eta^3} =& 1 + \delta T_6,\\
\frac {T_8}{\eta^6} =& 1 + 4 \delta T_6 +\delta T_8,\\
\frac {T_{66}}{\eta^5} =& 1 +2 \delta T_6 + \delta T_{66},\\
\frac {T_{44}}{\eta^4} =& 1 + \delta T_{44}.
\end{split}
\ee
The above equations should be understood as definitions of the symbols $\delta \eta$, $\delta T_6$, $\delta T_8$   $\delta T_{66}$ and $\delta T_{44}$ from the left-hand sides.  From eq. \eqref{exps} we see $\delta T_8$,  $\delta T_{66}$ and $\delta T_{44}$ are of order $1/N^3$ and $\delta T_6$ contains
both $1/N^2$ and $1/N^3$ terms.
To order $1/N^3$ the only mixed contribution is of the form  $\delta \eta \delta T_6$, while
the other corrections only contribute by their leading orders.  We thus obtain
\be
\eta_G-\eta^E &=& (-1)^{qE}(1+E \delta \eta)(T \delta T_6 + f_6 \delta T_8+ f_5 \delta T_{66}
+f_4 \delta T_{44}) \nn\\
&=& 
(-1)^{qE}(1+E T \delta \eta \delta T_6 + f_6 \delta T_8+ f_5 \delta T_{66}
+f_4 \delta T_{44} ) +O(1/N^4).
\ee
If we use the explicit expressions for the irreducible structures  given in eq. \eqref{exps}, we finally find
\be
\eta_G-\eta^E &= & (-1)^{Eq} \left(\frac{-8q^3}{N^2}\right)T+ (-1)^{Eq}\left [16ETq^5
 \right .  \nn\\ &&\left. +(-72T-80f_6-16f_5+16f_4)q^4  +32Tq^3\right ]\frac 1 {N^3}+ O(N^{-4}).
\ee
This fixes the form of $B(E,T, f_6, f_5, f_4)$ and proves eq. \eqref{eqn:etaDiffNcube}.

To calculate the correction to the Q-Hermite moments we sum over all intersection graphs:
\begin{equation}
\tilde M_{2p} - \tilde M_{2p}^{\text{QH}} =\sum_{i=1}^{(2p-1)!!}\left(\eta_{G_i}-\eta^{E_i}\right).
\end{equation}
It is now clear that  the new objects to be enumerated at order $1/N^3$ are $(-1)^{Eq}ET$ and $(-1)^{Eq}(5f_6+f_5-f_4)$. We will discuss their enumerations for the even $q$ and the odd $q$ cases in the following sections.

\subsubsection{Even $q$ case}

In the same way that $\tilde M_{2p}$ to order $1/N^3$ is determined by $\tilde M_4,\; \tilde M_6$ and $\tilde M_8$,  the Q-Hermite moment $\tilde M_{2p}^{\text{QH}}$ to order $1/N^3$ is determined by $\tilde M_4^{\text{QH}}, \tilde M_6^{\text{QH}}$ and $\tilde M_8^{\text{QH}}$. These Q-Hermite moments can be calculated most easily calculated from the definition
\be
\tilde M_{2p}^{\text{QH}} = \sum_{i=1}^{(2p-1)!!} \eta^{E_i},
\label{qhmoment}
\ee
the large $N$ expansion of $\eta$ in eq. \eqref{exps}, and the multiplicites (tables \ref{tab:6thMomIntersecGraphs} and \ref{tab:8thMomIntersecGraphs}). Alternatively, the sum in \eref{qhmoment} may be evaluated by the Riordan-Touchard
formula \cite{GG-J-V} where the multiplicies have been expressed in terms of binomial factors.
We find the moments:
\begin{equation}
\begin{split}
\tilde M_4^{\text{QH}} = &3-\frac{2 q^2}{N}+\frac{2 q^2(q-1)^2 }{N^2}-\frac{2 q^2 (q-1)^2 (2 q^2-8 q+5)}{3 N^3}+O\left(\frac{1}{N^4}\right),\\
\tilde M_6^{\text{QH}} =&15-\frac{30q^2}{N}+\frac{6 q^2 \left(9 q^2-10 q+5\right)}{N^2}\\
&-\frac{2 q^2 \left(38 q^4-108 q^3+139 q^2-90 q+25\right)}{N^3}+O(N^{-4}),\\
\tilde M_8^{\text{QH}} =&105-\frac{420q^2}{N}+\frac{28 q^2 \left(44 q^2-30 q+15\right)}{N^2}\\
&-\frac{4 q^2 \left(716 q^4-1232 q^3+1211 q^2-630 q+175\right)}{N^3}+O(N^{-4}). 
\end{split}
\end{equation}
The relation between cumulants and moments was given 
 in eq. \eqref{eqn:cumuNcubeEvenq}. Using the above Q-Hermite  moments, we get the following Q-Hermite cumulants:
\begin{equation}\label{eqn:QHcumuNcubeEvenq}
\begin{split}
  \tilde \kappa_4^{\text{QH}}
=&-\frac{2 q^2}{N}+\frac{2 q^2(q-1)^2 }{N^2}-\frac{2 q^2 (q-1)^2 (2 q^2-8 q+5)}{3 N^3}+O\left(N^{-4}\right),\\
\tilde \kappa_6^{\text{QH}}=
&\frac{24q^4}{N^2}-\frac{8 q^4 \left(7 q^2-12 q+6\right)}{N^3}+O\left(N^{-4}\right),\\
\tilde \kappa_8^{\text{QH}}=
&-\frac{736q^6}{N^3}+O\left(N^{-4}\right).
\end{split}
\end{equation}
The cumulants up to eighth order determine the $p$-dependence of all moments to order $1/N^3$ (see section
\ref{sec:3.2.1}).
The $p$-dependence of the moments is thus obtained from
eq. \eqref{eqn:m2pInCumu}:
\be
  \label{eqn:QHm2pNcubeEvenq}
\frac{\tilde M_{2p}^{\text{QH}}}{(2p-1)!!} &=&   1-\frac{2}{3} \binom{p}{2} \frac {q^2}{N}
  +\left[\binom{p}{2}\left(\frac{2}{3}q^2(q-1)^2\right)+\binom{p}{3}\frac{8}{5}q^4+\binom{p}{4}\frac{4}{3}q^4\right]\frac{1}{N^2}\nn\\
  &&-\left[\binom{p}{2}\left(\frac{2}{9}q^2(q-1)^2(2q^2-8q+5)\right)
+\binom{p}{3}\left(\frac{8}{15}  q^4 \left(7 q^2-12 q+6\right)\right)\right.\nn\\
&&\quad\ \ +\binom{p}{4} \left(\frac{8}{105} q^4 \left(127 q^2-70 q+35\right)\right)+ \binom{p}{5}\frac{32}{3}q^6\nn\\
&&\quad\ \ \left.+\binom{p}{6}\frac{40}{9}q^6\right]\frac{1}{N^3}+O\left(\frac{1}{N^4}\right).
  \ee
  We now use this result to derive enumerative identities for intersect on graphs.
  From the large $N$ expansion of $\eta$ in \eqref{exps} we obtain
\begin{equation}\label{eqn:etaElargeNcube}
\begin{split}
(-1)^{qE}\eta^E=&1-\frac{2 E q^2}{N}+\frac{2 E^2 q^4-4 E q^3+2 E q^2}{N^2}\\
&-\frac{4 E^3 q^6-24E^2 q^5+12 E^2 q^4+34E q^4-36 E q^3+10 E q^2}{3 N^3}+O\left(\frac{1}{N^4}\right).
\end{split}
\end{equation}
Matching the coefficients of powers of $q$ and $ 1/N$
in eqs. \eqref{eqn:QHm2pNcubeEvenq} and \eqref{qhmoment},
and this already gives several enumerative identities for $E$:
\begin{align}
\sum\limits_{i=1}^{(2p-1)!!}{E_i}
= &\frac{(2p-1)!!}{3}\binom{p}{2}, \label{eqn:totEdge}\\
\sum\limits_{i=1}^{(2p-1)!!}E_i^2
=&\frac{(2p-1)!!}{90}\binom{p}{2}(5p^2-p+12),\label{eqn:quadraticEdge}\\
\sum\limits_{i=1}^{(2p-1)!!}E_i^3=&\frac{(2p-1)!!}{3780}\binom{p}{2}\left(35 p^4+14 p^3+235 p^2-188 p+24\right).\label{eqn:cubicEdge}
\end{align}
In fact the above three identities are already known in mathematics literature \cite{flajolet2000}. However the proof therein is based on an analytical-combinatorial method quite distinct from ours. 

Now we proceed to calculate $\tilde M_{2p} - \tilde M_{2p}^{\text{QH}}$ for even $q$ from  eqs. \eqref{eqn:m2pNcubeEvenq} and \eqref{eqn:QHm2pNcubeEvenq}: 
\be
  \frac{\tilde M_{2p} - \tilde M_{2p}^{\text{QH}}}{(2p-1)!!}&=&-\frac 8 {15} \binom p 3 q^3 \frac{1}{N^2}+\left[\frac{8} {315}\binom{p}{3}( 7 p^2+ 5p+48)q^5 \right.
    \nn\\&&\left.+ \left(-\frac{24 }{5}\binom p 3-\frac{16}{15}\binom p 4\right) q^4 +\frac{32}{15}\binom p 3 q^3 \right]\frac{1}{N^3}+O(N^{-4}).
\ee
  Matching the coefficients of the expansion in $1/N$ and $q$  with
graph-theoretic result
eq. \eqref{eqn:etaDiffNcube}, we obtain the following enumerative identities:
\begin{align}
 \sum\limits_{i=1}^{(2p-1)!!}T_i &= \frac{(2p-1)!!}{15}\binom{p}{3},\label{eqn:totT} \\
  \sum_{i}E_iT_i&=\frac{(2p-1)!!}{630}\binom{p}{3}( 7 p^2+ 5p+48),\label{eqn:totET}\\
    \sum\limits_{i=1}^{(2p-1)!!}(5f_{6i}+f_{5i}-f_{4i}) &= \frac{(2p-1)!!}{15}\binom{p}{4},\label{eqn:totFourPoint}
\end{align}
where $f_{6i}, f_{5i}, f_{4i}$ are the numbers of  the three four-point structures in the intersection graph $G_i$. These are the truly new identities that could not have been obtained by the analytical-combinatorial method of \cite{flajolet2000}, since our identities involve higher structures such as triangles and four-point structures.

\subsubsection{Odd $q$ case}

For odd $q$ the SYK Hamiltonian $H$ is the supercharge of a supersymmetric theory with Hamiltonian
given by ${\cal H} = H^\dagger H$. The odd moments of this Hamiltonian do not vanish which changes
the relation between moments and cumulants.

Also for odd $q$ we start with the expansion of the moments of $H$ to order $1/N^3$. 
We repeat the same calculation as in the even  $q$ case, we easily obtain the
Q-Hermite moments up to eighth order
\begin{equation}
\begin{split}
\tilde M_4^{\text{QH}} = &1+\frac{2 q^2}{N} -2q^2 (q-1)^2 \frac{1}{N^2}+\frac{2}{3}q^2  (q-1)^2 \left(2 q^2-8 q+5\right) \frac{1}{N^3}+O\left(N^{-4}\right),\\
\tilde M_6^{\text{QH}} =&1+\frac{6q^2}{N} -\frac{6 (q-1)^2 q^2}{N^2}+\frac{2 q^2 \left(6 q^4-12 q^3+23 q^2-18 q+5\right)}{N^3}+O(N^{-4}),\\
\tilde M_8^{\text{QH}} =&1+\frac{12q^2}{N}+\frac{12 q^2 (2 q-1)}{N^2}-\frac{4 q^2(q-1)^2  (2 q-1) (2 q+5)}{N^3}+O(N^{-4}). 
\end{split}
\end{equation}
Using that the moments of ${\cal H}$ of order $p$ are equal to the moments of $H$ of order $2p$
we find the Q-Hermite cumulants (see eq. \eqref{eqn:cumulargeNoddq}):
\be
\tilde \kappa_2^{\text{QH}} &=&  \tilde M_4^{\text{QH}}-1 
=\frac{2 q^2}{N} -2q^2 (q-1)^2 \frac{1}{N^2}+\frac{2}{3} q^2(q-1)^2  \left(2 q^2-8 q+5\right) \frac{1}{N^3}+O\left(N^{-4}\right),\nn\\
\tilde \kappa_3^{\text{QH}} &=& \tilde M_6^{\text{QH}} -3 \tilde M_4^{\text{QH}} +2
= \frac{8 q^6}{N^3}+O\left(N^{-4}\right),\\
\tilde \kappa_4^{\text{QH}} &=& \tilde M_8^{\text{QH}} -4\tilde M_6^{\text{QH}}-3\left(\tilde M_4^{\text{QH}}\right)^2+12\tilde M_4^{\text{QH}}-6 
=-\frac{32q^6}{N^3}+O\left(N^{-4}\right).\nn
\ee
These cumulants determine the $p$-dependence of all moments up to order $1/N^3$ as is given in
eq.  \eqref{eqn:m2pInCumuOddq}. After substitution we obtain 
\be
\tilde M_{2p}^{\text{QH}}&=&1+2q^2\binom p 2 \frac{1}{N}+\left[\binom p 4 12q^4 -\binom p 2  2 (q-1)^2 q^2\right]\frac{1}{N^2} \nn\\
&&+\left[\binom p 2 \frac{2}{3}  q^2(q-1)^2 \left(2 q^2-8 q+5\right)+\binom p 3 8q^6  - \binom p 4 8q^4 \left(7 q^2-6 q+3\right)+\binom p 6 120q^6\right]\frac{1}{N^3}\nn\\&&+O\left(\frac{1}{N^4}\right).
\ee
Again, given  $\tilde M_{2p}^{\text{QH}} = \sum_i \eta^{E_i}$  and the form of $\eta$ in eq. \eqref{eqn:etaElargeNcube}, we have the following enumerative identities by matching the coefficients of the expansion in $1/N$ and $q$:
\begin{align}
\sum\limits_{i=1}^{(2p-1)!!}(-1)^{{E_i}}{E_i} &= -\binom{p}{2}, \label{eqn:gradedtotEdge}\\
\sum\limits_{i=1}^{(2p-1)!!}(-1)^{{E_i}}E_i^2&=\frac{1}{2}\binom{p}{2}(p-1)(p-4)
,\label{eqn:gradedQuadraticEdge}\\
\sum\limits_{i=1}^{(2p-1)!!} (-1)^{{E_i}}E_i^3&=-\frac{1}{4}\binom{p}{2} \left(p^4-14 p^3+57 p^2-76 p+24\right). \label{eqn:gradedCubicEdge}
\end{align}
These edge identities could also have been obtained by the analytical-combinatorial approach used in \cite{flajolet2000}. Now we calculate the difference between $\tilde{M}_{2p}$ and $\tilde{M}_{2p}^{\text{QH}}$:
\begin{equation}
\begin{split}
\tilde{M}_{2p}-\tilde{M}_{2p}^{\text{QH}}=&8q^3\binom p 3 \frac{1}{N^2}+\left[8\binom p 3 p(p-5)q^5+ \left(72\binom p 3 +16 \binom p 4\right)q^4-32\binom p 3 q^3\right]\frac{1}{N^3}\\
 &+O\left(\frac{1}{N^4}\right).
 \end{split}
\end{equation}
By matching with eq. \eqref{eqn:etaDiffNcube}, we obtain the following graded identities:
\begin{align}
\sum\limits_{i=1}^{(2p-1)!!} (-1)^{E_i}T_i &= - \binom{p}{3}, \label{eqn:gradedT}\\
\sum\limits_{i=1}^{(2p-1)!!} (-1)^{E_i}E_iT_i&=\frac{1}{2}\binom{p}{3}p( p-5),\label{eqn:gradedET}\\
\sum\limits_{i=1}^{(2p-1)!!}  (-1)^{E_i} (5f_{6i}+f_{5i}-f_{4i}) &= -\binom{p}{4}.\label{eqn:gradedFourPoint}
\end{align}
These identities could not have been obtained by using the method in \cite{flajolet2000}.

Finally we discuss in what sense the method presented in this paper is more
powerful than the method  developed in \cite{GG-J-V} to compute moments of the SYK model. The idea there is to solve for $\tilde M_{2p}-\tilde M_{2p}^{\text{QH}}$ for $q=1$ and $q=2$ models, where moments can be evaluated exactly,
and then do a matching with $\sum(\eta_{G_i}-\eta^{E_i})$ expressed in terms of graph-theoretic objects.
One can get a flavor of this method from appendix \ref{sec:consistency}.
The old method works very well to order $1/N^2$, however it becomes problematic
starting from order $1/N^3$. If we look at eq. \eqref{eqn:etaDiffNcube}, we see that at order $1/N^3$, on top of triangles, there are two new types of structures that need to be enumerated:
$(-1)^{Eq}ET$ for the $q^5$ term and $(-1)^{Eq}(5f_6+f_5-f_4)$
for the $q^4$ term. If we compute the moments for $q=1$ and $q=2$ models, at best we can obtain the enumeration of a linear combination of the two above-mentioned new structures, that is, $(-1)^{E}(ET-5f_6-f_5+f_4)$ for $q=1$ and $2ET-5f_6-f_5+f_4$ for $q=2$. However to recover the full $q$ dependence, we need separate enumerations of
$(-1)^{Eq}ET$ and $(-1)^{Eq}(5f_6+f_5-f_4)$. The method presented in this paper faces no such difficulty since the full $q$ dependence is retained at every stage of the calculation, and thus is capable of enumerating the two new structures separately. Nevertheless, the old method provides an independent consistency check of the results obtained in the present paper, which will be demonstrated in appendix \ref{sec:consistency}.

\section{Conclusions and outlook}
We have established the relation between the $1/N$ expansion of SYK moments and the $1/N$ expansion of the
high temperature expansion coefficients of the free energy.
It turns out that to compute the high temperature expansion of the free energy
to a certain order in $\beta$ and $1/N$, we only need to compute a finite number of moments to an appropriate order in $1/N$.
In particular, we have found in the thermodynamic limit the coefficients of order $\beta^6$
and $\beta^8$ for arbitrary $q$.
 The leading order $1/q^2$ results as well as the $1/q^3$ results are in agreement
with a large $N$ calculation of the path integral formulation of the SYK model, while the
$1/q^4$ correction was not calculated before and the exact result for the coefficients
reproduces the analytical result for $q=2$ obtained in a completely different way.

This relation also allows for calculations of moments to higher order in $1/N$, and
we have pushed the calculation to order $1/N^3$.
Surprisingly, we found that to a given order in $1/N$, all moments are determined
by a finite number of moments. This also explains the $p$-dependence of the moments
$\tilde M_{2p}$ obtained in previous work \cite{GG-J-V}. One important consequence of this is
that the SYK model generates elegant enumeration identities, at each order
in $1/N$, as discussed in section \ref{sec:enumeration}.
We tabulate some of them in table \ref{tab:enumID}. To the best of our knowledge,
it seems that only the identities at order $1/N$ are present in mathematical
literature \cite{flajolet2000}, and our study of the SYK model suggests they are only the
first layer of a hierarchy of identities. 
\renewcommand{\arraystretch}{2}
\begin{table}
 \begin{center}
\begin{tabular}{|c|c|c|}
\hline  & Non-SUSY& SUSY\\
\hline $1/N$&$\sum {E_i}
= \frac{(2p-1)!!}{3}\binom{p}{2}$ &$\sum (-1)^{{E_i}}{E_i} = -\binom{p}{2}$\\
 $1/N^2$& $\sum T_i = \frac{(2p-1)!!}{15}\binom{p}{3}$ &$\sum (-1)^{E_i}T_i = - \binom{p}{3}$\\
$1/N^3$& $\sum(5f_{6i}+f_{5i}-f_{4i}) = \frac{(2p-1)!!}{15}\binom{p}{4}$ &$\sum (-1)^{E_i} (5f_{6i}+f_{5i}-f_{4i}) = -\binom{p}{4}$\\
\hline
\end{tabular}
\end{center}
\caption{Some of the enumerative identities for intersection graphs generated at each order of $1/N$. All summation symbols run from $i=1$ to $i=(2p-1)!!$.}\label{tab:enumID}
\end{table}
\renewcommand{\arraystretch}{1}

It is clear that the procedure we presented can be extended to even higher orders in $1/N$, and the most computationally burdensome part is to calculate and
expand $\eta_G$ for the irreducible structures, for which we have a general and practical formula \cite{GG-J-V}.
For the large $N$ expansions of generic systems, there is often no powerful simplifying principle
that allows the calculation to be pushed to higher and higher order easily.
For SYK model, although the large $N$ coefficients of the moments are still somewhat complicated, the enumeration identities generated at each order of $1/N$ (table \ref{tab:enumID}) are incredibly simple. Does this simplicity of enumeration persist to higher and higher orders? If it does, does it imply there is
something we can say about the moments to all orders in $1/N$ instead of calculating them order by order?
We hope to clarify these questions in future studies.
Another issue is that we are in a somewhat awkward situation that on the one hand we have
the Q-Hermite expression for the spectral density \cite{GG-J-V},
which is a resummed approximation that is very accurate and gives the
leading order free energy for all temperatures, but Q-Hermite moments are only $1/N$-exact;
on the other hand, we have
expressions
for the moments that are exact to order $1/N^3$, but they only give the
high temperature information
of the free energy. It would be desirable to find an improved resummed expression
for the spectral density that goes beyond
the Q-Hermite approximation, which gives a free energy that is both $1/N^3$-exact and accurate at all temperatures.

\section{Acknowledgments}
{Mario Kieburg is thanked for pointing out the analytical
  result for $q=1$. Gerald Dunne is acknowledged for discussions on
  large order expansions and
  Antonio  Garc\'ia-Garc\'ia is thanked for a critical reading of the manuscript.
  Y.J. and J.V. acknowledge partial support from  U.S. DOE Grant  No. DE-FAG-88FR40388.
  After submission of the paper, an interesting preprint appeared
  \cite{Berkooz:2018qkz} that also
  uses chord diagrams to calculate a class of moments of the SYK Hamiltonian,
  but otherwise  has no significant overlap with the present work.
}

 \appendix

\section{High temperature expansion of free energy from \cite{Tarnopolsky}} \label{append:tarnopolsky}
In section \ref{sec:evenqFreeE} we claimed that our result for the free energy for even $q$ reproduces an independent calculation \cite{Tarnopolsky} to order $1/q^3$, and in this appendix we justify this claim.

In \cite{Tarnopolsky} the large $q$ expansion for the extensive (leading in $1/N$) part of the free energy is given:
\begin{equation}\label{eqn:tarnopolskyFreeEnergy}
\frac{-\beta F}{N} =\frac{1}{2}\log 2 +\frac{1}{q^2} \pi u \left(\tan \frac{\pi u}{2}-\frac{\pi u}{4}\right)+\frac{1}{q^3}\pi u \left[\pi u-2\tan \frac{\pi u}{2}\left(1-\frac{\pi^2 u^2}{12}\right)\right],
\end{equation}
where 
\begin{equation}
\frac{\pi u}{\cos \frac{\pi u}{2}}=\beta J^2 2^{1-q}q.
\end{equation}
If we set $J^2=2^{q-1}/q$ as we did in section \ref{sec:evenqFreeE}, we have simply 
\begin{equation}\label{eqn:tarnopolskyVariableu}
\frac{\pi u}{\cos \frac{\pi u}{2}}=\beta.
\end{equation}
Then by iterating eq. \eqref{eqn:tarnopolskyVariableu} we obtain
\begin{equation}
  u=\frac{1}{\pi}\left(\beta-\frac{1}{3!}\frac{3}{4}\beta^3+\frac{1}{5!}\frac{65}{16}\beta^5-\frac{1}{7!}\frac{3787}{64}\beta^7\right)+O(\beta^9).
  \label{uexp}
\end{equation}
Using this high temperature expansion of $u$, we obtain for the
free energy, eq. \eqref{eqn:tarnopolskyFreeEnergy},  
\begin{equation}
\frac{-\beta F}{N} = \frac{\log 2}{2}+\frac{1}{2!}\frac{\beta^2}{2q^2}-\frac{1}{4!}\frac{\beta^4}{2 q^2}+\frac{1}{6!}\frac{\beta^6 (3 q-1)}{
   q^3}+\frac{1}{8!}\frac{\beta^8 (36-46 q)}{ q^3}+O\left(\beta^{10}\right).
\end{equation}
This is consistent with our result eq. \eqref{eqn:freeEnergyHighTempNcubeEvenq} to order $1/q^3$ at leading order in $1/N$.

 The advantage of the result of \cite{Tarnopolsky} is that it is valid at all temperatures, however our result, albeit only valid at high temperatures, gives the expansion to even higher orders in $1/q$ and $1/N$.

 As a side remark, the coefficients of the high temperature expansion of
 $u$ are known analytically. If we write
 \be
 u = \frac 1\pi\sum_{n=0}^\infty \frac{(-1)^n}{(2n+1)!2^{2n}} a_n \beta^{2n+1},
   \ee
   then  $a_n$ are given by
   \be
   a_n= \frac 1{2^{2n+1}} \sum_{k=0}^{2n+1} {2n+1\choose k}(2k-2n-1)^{2n}.
   \ee
   This is a convergent series with a convergence radius of $\beta_c \approx 1.33$.
   The numbers $a_n$ are also the number of labeled rooted trees on $2n+1$
   nodes with each node having an even number of children \cite{oeis}.
   Since there is no singularity on the positive real axis, we do not expect a
   phase transition as $\beta$ varies.
   
 \section{High temperature free energy to order $1/N^3$}\label{appen:freeEnergyNcube}
 In the main text, we have only displayed the leading order of the large $N$ expansions of the free energies in eqs. \eqref{eqn:freeEnergyHighTempNcubeEvenq} and \eqref{eqn:freeEnergyHighTempNcube}. In this appendix we display the large $N$ series of free energies to order $1/N^3$, and again we adopt the normalization convention $J^2 = 2^{q-1}/q$.
 \subsection{Even $q$ case}
 \begin{equation}
\begin{split}
\frac{-\beta F}{N}= &\frac 12 \log 2 +\frac 1{2q^2}  \frac{\beta ^2}{2!} -\frac{1}{2 q^2 }  \frac{\beta^4}{4!}
+ \frac{(3q-1)}{q^3 }\frac{\beta^6}{6!} - \frac{(46q^2-36q+7)}{q^4 } \frac{\beta^8}{8!} +O(\beta^{10})\\
&+\frac{1}{N}\left[-\frac{(q-1)}{4q}\frac{\beta^2}{2!}+\frac{(q-1) (2 q-1)}{2 q^2}\frac{\beta^4}{4!}-\frac{(q-1) \left(23 q^2-25 q+8\right)}{2 q^3}\frac{\beta ^6}{6!}\right.\\
&\qquad\quad \left.+\frac{2  (q-1) \left(46 q^2-36 q+7\right)}{q^3}\frac{\beta ^8}{8!}+O(\beta^{10})\right]\\
&+\frac{1}{N^2}\left[\frac{(q-2) (q-1) (3 q-1)}{ 48 q}\frac{\beta ^2}{2!} -\frac{ (q-1) \left(13 q^3-37 q^2+33 q-10\right)}{12 q^2}\frac{\beta ^4}{4!} \right. \\
&\qquad\quad\left.+\frac{ (q-1) \left(111 q^3-264 q^2+199 q-50\right)}{8 q^2} \frac{\beta^6}{6!}\right.\\
&\qquad\quad \left.-\frac{(q-1) ( 6 q^2- 8 q +1) (46 q^2- 36 q+7)}{
 3 q^3}\frac{\beta^8}{8!}+O(\beta^{10})\right]\\
 &+\frac{1}{N^3}\left[-\frac{ (q-3) (q-2) (q-1)^2}{96}\frac{\beta^2}{2!}+\frac{ (q-1)^3 \left(8 q^2-23 q+11\right)}{12 q}\frac{\beta ^4}{4!}\right.\\
 &\qquad\quad \left. -\frac{ (q-1)^2 \left(153 q^4-452 q^3+437 q^2-158 q+16\right)}{16 q^2}\frac{\beta ^6}{6!}\right.\\
 &\qquad\quad \left. +\frac{ (q-1)^2 (2 q-3) (2 q-1) \left(46 q^2-36 q+7\right)}{3 q^2}\frac{\beta ^8}{8!}+O(\beta^{10})\right]+O(1/N^4).\\
\end{split}
\end{equation}
\subsection{Odd $q$ case}
\begin{equation}
\begin{split}
\frac{-\beta F}{N}= &\frac 12 \log 2-\frac 1{2 q^2}  \beta  +\frac{1}{2 q^2 } \frac{ \beta^2}{2!}
- \frac{1}{ q^3 }\frac{\beta^3}{3!} + \frac{-(2q^2-4q-1)}{ q^4} \frac{\beta^4}{4!} +O(\beta^{5})\\
&+\frac{1}{N} \left[\frac{q-1}{4 q}\beta-\frac{(q-1) (2 q-1)}{2 q^2}\frac{\beta ^2}{2!}-\frac{(q-1) \left(2 q^2-13
   q+8\right)}{2 q^3}\frac{\beta ^3}{3!}\right.\\
   &\qquad\quad\left.+\frac{2 (q-1) \left(2 q^2-4 q-1\right)}{q^3}\frac{\beta ^4}{4!}+O(\beta^{5})\right]\\
   &+ \frac{1}{N^2}\left[-\frac{(q-2) (q-1) (3 q-1)}{48 q}\beta+\frac{(q-1) \left(13 q^3-37 q^2+33
   q-10\right)}{12 q^2}\frac{\beta ^2}{2!}\right.\\
   &\qquad\quad \left. +\frac{(q-1) \left(12 q^3-81 q^2+121 q-50\right)}{8
   q^2}\frac{\beta ^3}{3!}\right.\\
   &\qquad\quad\left. -\frac{(q-1) \left(2 q^2-4 q-1\right) \left(6 q^2-8 q+1\right)}{3
   q^3}\frac{\beta ^4}{4!}+O(\beta^5)\right]\\
   &+\frac{1}{N^3}\left[\frac{(q-3) (q-2) (q-1)^2}{96}\beta -\frac{(q-1)^3 \left(8 q^2-23
   q+11\right)}{12 q}\frac{\beta ^2}{2!}\right.\\
   &\qquad\quad\left. -\frac{(q-1)^2 \left(18 q^4-125 q^3+227 q^2-134 q+16\right)}{16
   q^2}\frac{\beta ^3}{3!}\right.\\
   &\qquad\quad\left.+\frac{(q-1)^2 (2 q-3) (2 q-1) \left(2 q^2-4 q-1\right)}{3 q^2}\frac{\beta ^4}{4!}+O(\beta^5)\right]+O(1/N^4).
\end{split}
\end{equation}
\section{Consistency check by computing the $q=1$ and $q=2$ SYK models}\label{sec:consistency}
In the main text we showed that the $1/N^3$ coefficient of $\eta_G-\eta^E$ is 
\be\label{eqn:NcubeDiff}
\eta_G-\eta^E \cong \left[16ETq^5+(-72T-80f_6-16f_5+16f_4)q^4+32Tq^3\right]\frac{1}{N^3},
\ee
where the notation ``$\cong$'' means everything but the $1/N^3$ term is omitted. We will calculate this quantity summed over all intersection graphs for $q=1$ and $q=2$. By Wick's theorem, we need to compute $\tilde M_{2p}-\tilde M^{\text{QH}}_{2p}$.

\subsection{$q=1$ case} 

As discussed in detail in \cite{GG-J-V}, for $q=1$, $H^{2p}= \left(\sum_{\alpha=1}^N J_\alpha^2\right)^{p}\mathbb{1}$, and because $J_\alpha$ is Gaussian distributed, $\left(\sum_{\alpha=1}^N J_\alpha^2\right)^{p}$ follows a $\chi^2$ distribution. Hence for the $q=1$ SYK model we have 
\be
\tilde M_{2p} = \frac{\Gamma\left(\frac N2 +p\right)}{ \left(\frac N2\right)^p\Gamma\left(\frac N2\right)}
\cong \frac{(p-1)^2 p^2 \left(p^2-5 p+6\right)}{6 N^3},
  \ee
  where again we have only kept the $1/N^3$ term. 

  To evaluate $\tilde M^{\text{QH}}$, we first expand $\eta^E$ to $1/N^3$.
For $q=1$  the first equation of  \eqref{exps} simplifies to
\be
\label{eqn:etaToNcube}
  (-1)^q\eta=1-\frac{2 }{N}
\ee
and the binomial expansion gives the $1/N^3$ correction
\be
\eta^E\cong -(-1)^{E}\frac{4E(E-1)(E-2)}{3 N^3}.
\ee
The enumeration of $E(E-1)(E-2)$ for odd $q$ was worked out by our method in
eqs. \eqref{eqn:gradedtotEdge}, \eqref{eqn:gradedQuadraticEdge} and
\eqref{eqn:gradedCubicEdge}, but as mentioned it can also be obtained by a completely independent method. Using eq. \eqref{eqn:gradedCubicEdge}, we obtain 
\begin{equation}
\tilde M^{\text{QH}}_{2p}\cong \binom{p}{3}(p^3 -12p^2+39 p-28)\frac{1}{N^3}
\end{equation}
Hence we find
\begin{equation}
\tilde M_{2p}-\tilde M^{\text{QH}}_{2p}\cong \frac{2}{3}p (p-1)^2  \left(2 p^2-11 p+14\right)\frac{1}{N^3}.
\end{equation}
This result is consistent with eqs. \eqref{eqn:gradedT}, \eqref{eqn:gradedET}, \eqref{eqn:gradedFourPoint} and \eqref{eqn:NcubeDiff}.

\subsection{$q=2$ case} 
As explained in detail in \cite{GG-J-V}, for $q=2$ we have 
\be\label{eqn:q=2moments}
\tilde M_{2p}=\left.\left\langle\left(\sum_{k=1}^{N/2}x_k\right)^{2p}\right\rangle \middle/ \left(\frac{N}{2} \langle x_1^2\rangle\right)^p \right.
\ee
where the brackets $\langle\cdots \rangle$ on the right-hand side denote the ensemble average with the  probability distribution \cite{mehta2004,gross2017},
\be
P(x_1,\ldots,x_{N/2}) \prod_{l=1}^{N/2}dx_l= c e^{-\sum_k x_k^2}\prod_{i<j}(x_i^2-x_j^2)^2\prod_{l=1}^{N/2}dx_l,
\ee
and the constant $c$ normalizes the total probability to one. We can do a multinomial expansion for the right-hand side of eq. \eqref{eqn:q=2moments}:
\be
\label{eqn:q2MomPartition}
\left\langle\left(\sum_{k=1}^{N/2}x_k\right)^{2p}\right\rangle= \sum_{m_1+\cdots +m_{N/2}=p}\frac{(2p)!}{(2m_1)!(2m_2)!\cdots(2m_{N/2})!}\left\langle x_1^{2m_1}x_2^{2m_2}\cdots x_{N/2}^{2m_{N/2}}\right\rangle.\ \ 
\ee
Following the argument laid out in \cite{GG-J-V}, we conclude only the following terms contribute to $\tilde M_{2p}$ to $1/N^3$ order: 
\be
\begin{split}
\tilde M_{2p}=&\left(\frac{N}{2}\right)^{-p}\binom{N/2}{p} \frac{(2p)!}{2^p}\frac{W_1}{W_0^p}  +\left(\frac{N}{2}\right)^{-p}\binom{N/2}{p-1}\binom{p-1}{1} \frac{(2p)!}{2^{p-2}4!}\frac{W_2}{W_0^p} \\
&+\left(\frac{N}{2}\right)^{-p}\binom{N/2}{p-2}\left[ \binom{p-2}{2}\frac{(2p)!}{2^{p-4}4!4!}\frac{W_3}{W_0^p}+\binom{p-2}{1}\frac{(2p)!}{2^{p-3}6!}\frac{W_4}{W_0^p}
\right]\\
&+\left(\frac{N}{2}\right)^{-p}\binom{N/2}{p-3}\left[\binom{p-3}{3}\frac{(2p)!}{2^{p-6}(4!)^3}\frac{W_5}{W_0^p}+2\binom{p-3}{2}\frac{(2p)!}{2^{p-5}6!4!}\frac{W_6}{W_0^p}+\right.\\
&\left.\qquad\qquad\qquad\qquad\quad\ \binom{p-3}{1}\frac{(2p)!}{2^{p-4}8!}\frac{W_7}{W_0^p}\right]+O(1/N^4),
\end{split}
\ee
where 
\be\label{eqn:selberg}
\begin{split}
&W_0 := \langle x_1^2\rangle, \\
&W_1 := \left\langle x_1^{2}x_2^{2}\cdots x_{p}^{2}\right\rangle, \\
&W_2 := \left\langle x_1^{4}x_2^{2}\cdots x_{p-1}^{2}\right\rangle, \\
&W_3 :=\left\langle x_1^{4}x_2^{4}x_3^2\cdots x_{p-2}^{2}\right\rangle, \\
&W_4 :=\left\langle x_1^{6}x_2^{2}\cdots x_{p-2}^{2}\right\rangle, \\
&W_5 :=\left\langle x_1^{4}x_2^{4}x_3^4x_4^2\cdots x_{p-3}^{2}\right\rangle, \\
&W_6 :=\left\langle x_1^{6}x_2^{4}x_3^2\cdots x_{p-3}^{2}\right\rangle, \\
&W_7 :=\left\langle x_1^{8}x_2^{2}\cdots x_{p-3}^{2}\right\rangle. \\
\end{split}
\ee
Note that all $W_i$ are of the form of a Selberg-like integral. Before evaluating $W_i$, we first  expand
the prefactors to $1/N^3$:
\be
\label{eqn:q2MomentIntermediate}
 \begin{split}
\tilde M_{2p}=&(2p-1)!!\left\{\left[1-2\binom{p}{2}\frac{1}{N}+(3p-1)\binom{p}{3}\frac{1}{N^2}-8\binom{p}{4}\binom{p}{2}\frac{1}{N^3}\right]\frac{W_1}{W_0^p}  \right.\\ 
&\qquad\qquad\quad+\left[\frac{2}{3}\binom{p}{2}\frac{1}{N}-2(p-1)\binom{p}{3}\frac{1}{N^2}+\frac{4}{3}(p-1)(3p-4)\binom{p}{4}\frac{1}{N^3}\right]\frac{W_2}{W_0^p} \\
&\qquad\qquad\quad +\left[ \frac{4}{3}\binom{p}{4}\frac{1}{N^2}-\frac{8}{3}\binom{p}{4}\binom{p-2}{2}\frac{1}{N^3}\right]\frac{W_3}{W_0^p} \\
 &\qquad\qquad\quad\left.+\left[\frac{4}{15}\binom{p}{3}\frac{1}{N^2}-\frac{16}{15}(p-2)\binom{p}{4}\frac{1}{N^3}\right]\frac{W_4}{W_0^p}+\frac{40}{9}\binom{p}{6}\frac{1}{N^3}\frac{W_5}{W_0^p} \right. \\
 &\left.\qquad\qquad\quad+\frac{16}{9}\binom{p}{5}\frac{1}{N^3}\frac{W_6}{W_0^p}+\frac{8}{105}\binom{p}{4}\frac{1}{N^3}\frac{W_7}{W_0^p}
\right\}+O\left(1/N^4\right).
\end{split}
\ee
We can work out the values of the $W_i$'s by employing a set of recursion relations for Selberg integrals developed in \cite{verbaarschot:1994gr} resulting in
\be
\begin{split}
&W_0 = \frac{N-1}{2}, \\
&W_1 = \prod_{k=0}^{p-1}\left(\frac{N}{2}-p+k+\frac{1}{2}\right), \\
&W_2 =\left(N-p+\frac{3}{2}\right)\prod_{k=0}^{p-2}\left(\frac{N}{2}-p+k+\frac{3}{2}\right), \\
&W_3 =\left(N-p+\frac{3}{2}\right)\left(N-p+\frac{5}{2}\right)\prod_{k=0}^{p-3}\left(\frac{N}{2}-p+k+\frac{5}{2}\right), \\
&W_4 =\left(N+\frac{1}{2}\right)\left(N-p+\frac{5}{2}\right)\prod_{k=0}^{p-3}\left(\frac{N}{2}-p+k+\frac{5}{2}\right)\\
&\qquad\quad+\left(\frac{N}{2}-p+2\right)\prod_{k=0}^{p-2}\left(\frac{N}{2}-p+k+\frac{3}{2}\right), \\
&W_5  =\prod_{k=0}^2\left(N-p+\frac{3}{2}+k\right)\prod_{l=0}^{p-4}\left(\frac{N}{2}-p+l+\frac{7}{2}\right), \\
&W_6 =\left(N-\frac{1}{2}\right)\prod_{k=0}^1\left(N-p+\frac{5}{2}+k\right)\prod_{l=0}^{p-4}\left(\frac{N}{2}-p+l+\frac{7}{2}\right)\\
&\qquad\quad +\left(\frac{N}{2}-p+3\right)\left(N-p+\frac{5}{2}\right)\prod_{k=0}^{p-3}\left(\frac{N}{2}-p+k+\frac{5}{2}\right), \\
&W_7 =\left(N+\frac{3}{2}\right)\left(N+\frac{1}{2}\right)\left(N-p+\frac{7}{2}\right)\prod_{l=0}^{p-4}\left(\frac{N}{2}-p+l+\frac{7}{2}\right)\\
&\qquad\quad +\left(N+\frac{3}{2}\right)\left(\frac{N}{2}-p+3\right)\prod_{k=0}^{p-3}\left(\frac{N}{2}-p+k+\frac{5}{2}\right)\\
&\quad\quad\ \ +(p-4)\prod_{k=0}^1\left(N-p+\frac{5}{2}+k\right)\prod_{l=0}^{p-4}\left(\frac{N}{2}-p+l+\frac{7}{2}\right)\\
&\qquad\quad +\left(N-2p+6\right)\left(N-p+\frac{5}{2}\right)\prod_{l=0}^{p-3}\left(\frac{N}{2}-p+l+\frac{5}{2}\right).
\end{split}
\ee
To relevant order, we have 
\be
\begin{split}
&\frac{W_1}{W_0^p} = 1-2\binom{p}{2}\frac{1}{N}+\frac{1}{3}\binom{p}{2}(3p^2-7p-4)\frac{1}{N^2}+\frac{1}{3}\binom{p}{2}(p^2-4p-2)(p^2-2p-1)\frac{1}{N^3}, \\
&\frac{W_2}{W_0^p}  =2-(2p^2-4p-1)\frac{1}{N}+\frac{1}{3}(3 p^4- 16 p^3+ 18 p^2+ 7 p -3 )\frac{1}{N^2}, \\
&\frac{W_3}{W_0^p}  =4-4(p^2-3p)\frac{1}{N}, \\
&\frac{W_4}{W_0^p} =5-(5p^2-17p+1)\frac{1}{N}, \\
&\frac{W_5}{W_0^p} =8, \\
&\frac{W_6}{W_0^p} =10, \\
&\frac{W_7}{W_0^p} =14.
\end{split}
\ee
For $q=2$ we finally arrive at
\be
\tilde M_{2p}\cong(2p-1)!!\left(-\frac{32 p^6}{81}+\frac{32 p^5}{27}-\frac{16 p^4}{405}-\frac{128 p^3}{135}-\frac{436
   p^2}{405}+\frac{172 p}{135}\right)\frac{1}{N^3},
\ee
where we have omitted all but the $1/N^3$ term. The results up to order $1/N^2$ can be found in \cite{GG-J-V}. 
To evaluate $\tilde M_{2p}^{\text {QH}}$, the $1/N$ expansion of $\eta$ in
the first equation of  \eqref{exps} simplifies for $q=2$ to
\be
\eta=1-\frac 8N+\frac 8{N^2} +\frac{8}{N^3}+ O(1/N^4),
\ee
and hence
\be\eta^E \cong -\frac{256E^3-576E^2+296E}{3 N^3} \ee
for $q=2$. 
Using eqs. \eqref{eqn:totEdge}, \eqref{eqn:quadraticEdge} and \eqref{eqn:cubicEdge}, we obtain
\begin{equation}
\tilde M_{2p}^{\text{QH}} \cong (2p-1)!!\left( -\frac{32 p^6}{81}+\frac{32 p^5}{135}+\frac{8048 p^4}{2835}-\frac{512
   p^3}{315}-\frac{14092 p^2}{2835}+\frac{740 p}{189}\right)\frac{1}{N^3}.
\end{equation}
Finally, 
\be
\tilde M_{2p}-\tilde M_{2p}^{\text{QH}}\cong (2p-1)!!\left(\frac{128 p^5}{135}-\frac{544 p^4}{189}+\frac{128 p^3}{189}+\frac{736
   p^2}{189}-\frac{832 p}{315}\right)\frac{1}{N^3}
\ee
which agrees with eqs. \eqref{eqn:totT},  \eqref{eqn:totET}, \eqref{eqn:totFourPoint} and \eqref{eqn:NcubeDiff}.

\bibliographystyle{jhep}
\bibliography{library-5-23}

\providecommand{\href}[2]{#2}\begingroup\raggedright\begin{thebibliography}{10}

\bibitem{GG-J-V}
A.~M. García-García, Y.~Jia and J.~J.~M. Verbaarschot, \emph{{Exact moments
  of the Sachdev-Ye-Kitaev model up to order $1/N^2$}},
  \href{https://arxiv.org/abs/1801.02696}{{\ttfamily 1801.02696}}.

\bibitem{Tarnopolsky}
G.~Tarnopolsky, \emph{{On large $q$ expansion in the Sachdev-Ye-Kitaev model}},
   \href{https://arxiv.org/abs/1801.06871}{{\ttfamily 1801.06871}}.

\bibitem{bethe1936}
H.~A. Bethe, \emph{An attempt to calculate the number of energy levels of a
  heavy nucleus}, \href{https://doi.org/10.1103/PhysRev.50.332}{\emph{Phys.
  Rev.} {\bfseries 50} (Aug, 1936) 332--341}.

\bibitem{Egidy:2005aw}
T.~von Egidy and D.~Bucurescu, \emph{{Systematics of nuclear level density
  parameters}}, \href{https://doi.org/10.1103/PhysRevC.72.044311,
  10.1103/PhysRevC.73.049901}{\emph{Phys. Rev.} {\bfseries C72} (2005) 044311}.

\bibitem{french1970}
J.~B. French and S.~S.~M. Wong, \emph{{Validity of random matrix theories for
  many-particle systems}},
  \href{https://doi.org/10.1016/0370-2693(70)90213-3}{\emph{Phys. Lett.}
  {\bfseries 33B} (1970) 449--452}.

\bibitem{french1971}
J.~B. French and S.~S.~M. Wong, \emph{{Some random-matrix level and spacing
  distributions for fixed-particle-rank interactions}},
  \href{https://doi.org/10.1016/0370-2693(71)90424-2}{\emph{Phys. Lett.}
  {\bfseries 35B} (1971) 5--7}.

\bibitem{bohigas1971}
O.~Bohigas and J.~Flores, \emph{{Two-body random hamiltonian and level
  density}}, \href{https://doi.org/10.1016/0370-2693(71)90598-3}{\emph{Phys.
  Lett.} {\bfseries 34B} (1971) 261--263}.

\bibitem{bohigas1971a}
O.~Bohigas and J.~Flores, \emph{{Spacing and individual eigenvalue
  distributions of two-body random hamiltonians}},
  \href{https://doi.org/10.1016/0370-2693(71)90399-6}{\emph{Phys. Lett.}
  {\bfseries 35B} (1971) 383--386}.

\bibitem{garcia2017}
A.~García-García and J.~Verbaarschot, \emph{{Analytical Spectral Density of
  the Sachdev-Ye-Kitaev Model at finite N}},
  \href{https://doi.org/10.1103/PhysRevD.96.066012}{\emph{Phys. Rev.}
  {\bfseries D96} (2017) 066012},
  [\href{https://arxiv.org/abs/1701.06593}{{\ttfamily 1701.06593}}].

\bibitem{brody1981}
T.~Brody, J.~Flores, J.~French, P.~Mello, A.~Pandey and S.~Wong,
  \emph{Random-matrix physics: spectrum and strength fluctuations},
  \href{https://doi.org/10.1103/RevModPhys.53.385}{\emph{Rev. Mod. Phys.}
  {\bfseries 53} (Jul, 1981) 385--479}.

\bibitem{kota2001}
V.~Kota, \emph{Embedded random matrix ensembles for complexity and chaos in
  finite interacting particle systems},
  \href{https://doi.org/http://dx.doi.org/10.1016/S0370-1573(00)00113-7}{\emph{Physics
  Reports} {\bfseries 347} (2001) 223 -- 288}.

\bibitem{gomez2011}
J.~Gomez, K.~Kar, V.~Kota, R.~Molina, A.~Relano and J.~Retamosa,
  \emph{Many-body quantum chaos: Recent developments and applications to
  nuclei},
  \href{https://doi.org/http://dx.doi.org/10.1016/j.physrep.2010.11.003}{\emph{Physics
  Reports} {\bfseries 499} (2011) 103 -- 226}.

\bibitem{kota2011a}
V.~K.~B. Kota, A.~Rela{\~n}o, J.~Retamosa and M.~Vyas, \emph{Thermalization in
  the two-body random ensemble}, {\emph{Journal of Statistical Mechanics:
  Theory and Experiment} {\bfseries 2011} (2011) P10028}.

\bibitem{kota2014}
V.~K.~B. Kota, \emph{Embedded random matrix ensembles in quantum physics},
  vol.~884.
\newblock Springer, 2014.

\bibitem{garcia2017a}
A.~Garc{\'\i}a-Garc{\'\i}a, A.~Romero-Berm{\'u}dez and M.~Tezuka,
  \emph{Stability of chaos in a generalised sachdev-ye-kitaev model},
  \href{https://arxiv.org/abs/1702.01738}{{\ttfamily 1702.01738}}.

\bibitem{Nosaka:2018iat}
T.~Nosaka, D.~Rosa and J.~Yoon, \emph{{Thouless time for mass-deformed SYK}},
  \href{https://arxiv.org/abs/1804.09934}{{\ttfamily 1804.09934}}.

\bibitem{Garcia-Garcia:2018ruf}
A.~M. García-García, Y.~Jia and J.~J.~M. Verbaarschot, \emph{{Universality
  and Thouless energy in the supersymmetric Sachdev-Ye-Kitaev Model}},
  \href{https://doi.org/10.1103/PhysRevD.97.106003}{\emph{Phys. Rev.}
  {\bfseries D97} (2018) 106003},
  [\href{https://arxiv.org/abs/1801.01071}{{\ttfamily 1801.01071}}].

\bibitem{Bohigas:1974zz}
O.~Bohigas, J.~Flores, J.~B. French, M.~J. Giannoni, P.~A. Mello and S.~S.~M.
  Wong, \emph{{Recent results on energy-level fluctuations}},
  \href{https://doi.org/10.1103/PhysRevC.10.1551}{\emph{Phys. Rev.} {\bfseries
  C10} (1974) 1551--1553}.

\bibitem{benet2001}
L.~Benet, T.~Rupp and H.~Weidenmuller, \emph{{Nonuniversal behavior of the k
  body embedded Gaussian unitary ensemble of random matrices}},
  \href{https://doi.org/10.1103/PhysRevLett.87.010601}{\emph{Phys. Rev. Lett.}
  {\bfseries 87} (2001) 010601},
  [\href{https://arxiv.org/abs/cond-mat/0010425}{{\ttfamily
  cond-mat/0010425}}].

\bibitem{Altland:2017eao}
A.~Altland and D.~Bagrets, \emph{{Quantum ergodicity in the SYK model}},
  \href{https://doi.org/10.1016/j.nuclphysb.2018.02.015}{\emph{Nucl. Phys.}
  {\bfseries B930} (2018) 45--68},
  [\href{https://arxiv.org/abs/1712.05073}{{\ttfamily 1712.05073}}].

\bibitem{benet2003}
L.~Benet and H.~A. Weidenmuller, \emph{{Review of the k body embedded ensembles
  of Gaussian random matrices}},
  \href{https://doi.org/10.1088/0305-4470/36/12/340}{\emph{J. Phys.} {\bfseries
  A36} (2003) 3569--3594},
  [\href{https://arxiv.org/abs/cond-mat/0207656}{{\ttfamily
  cond-mat/0207656}}].

\bibitem{borgonovi2016}
F.~{Borgonovi}, F.~M. {Izrailev}, L.~F. {Santos} and V.~G. {Zelevinsky},
  \emph{{Quantum chaos and thermalization in isolated systems of interacting
  particles}}, \href{https://doi.org/10.1016/j.physrep.2016.02.005}{\emph{Phys.
  Rep.} {\bfseries 626} (Apr., 2016) 1--58},
  [\href{https://arxiv.org/abs/1602.01874}{{\ttfamily 1602.01874}}].

\bibitem{sachdev1993}
S.~Sachdev and J.~Ye, \emph{{Gapless spin fluid ground state in a random,
  quantum Heisenberg magnet}},
  \href{https://doi.org/10.1103/PhysRevLett.70.3339}{\emph{Phys. Rev. Lett.}
  {\bfseries 70} (1993) 3339},
  [\href{https://arxiv.org/abs/cond-mat/9212030}{{\ttfamily
  cond-mat/9212030}}].

\bibitem{sachdev00}
A.~{Georges}, O.~{Parcollet} and S.~{Sachdev}, \emph{{Quantum fluctuations of a
  nearly critical Heisenberg spin glass}},
  \href{https://doi.org/10.1103/PhysRevB.63.134406}{\emph{Phys. Rev.}
  {\bfseries B63} (Apr., 2001) 134406},
  [\href{https://arxiv.org/abs/cond-mat/0009388}{{\ttfamily
  cond-mat/0009388}}].

\bibitem{Sachdev:2010um}
S.~Sachdev, \emph{{Holographic metals and the fractionalized Fermi liquid}},
  \href{https://doi.org/10.1103/PhysRevLett.105.151602}{\emph{Phys. Rev. Lett.}
  {\bfseries 105} (2010) 151602},
  [\href{https://arxiv.org/abs/1006.3794}{{\ttfamily 1006.3794}}].

\bibitem{Huang:2017nox}
Y.~Huang and Y.~Gu, \emph{{Eigenstate entanglement in the Sachdev-Ye-Kitaev
  model}},  \href{https://arxiv.org/abs/1709.09160}{{\ttfamily 1709.09160}}.

\bibitem{kitaev2015}
A.~Kitaev, \emph{A simple model of quantum holography}.
\newblock KITP strings seminar and Entanglement 2015 program,
  \url{http://online.kitp.ucsb.edu/online/entangled15/}, 2015.

\bibitem{maldacena2016}
J.~Maldacena and D.~Stanford, \emph{{Remarks on the Sachdev-Ye-Kitaev model}},
  \href{https://doi.org/10.1103/PhysRevD.94.106002}{\emph{Phys. Rev.}
  {\bfseries D94} (2016) 106002},
  [\href{https://arxiv.org/abs/1604.07818}{{\ttfamily 1604.07818}}].

\bibitem{jensen2016}
K.~Jensen, \emph{{Chaos in AdS$_2$ Holography}},
  \href{https://doi.org/10.1103/PhysRevLett.117.111601}{\emph{Phys. Rev. Lett.}
  {\bfseries 117} (2016) 111601},
  [\href{https://arxiv.org/abs/1605.06098}{{\ttfamily 1605.06098}}].

\bibitem{polchinski2016}
J.~Polchinski and V.~Rosenhaus, \emph{{The Spectrum in the Sachdev-Ye-Kitaev
  Model}}, \href{https://doi.org/10.1007/JHEP04(2016)001}{\emph{JHEP}
  {\bfseries 04} (2016) 001},
  [\href{https://arxiv.org/abs/1601.06768}{{\ttfamily 1601.06768}}].

\bibitem{polchinski2016a}
J.~Polchinski and V.~Rosenhaus, \emph{The spectrum in the sachdev-ye-kitaev
  model}, \href{https://doi.org/10.1007/JHEP04(2016)001}{\emph{Journal of High
  Energy Physics} {\bfseries 04} (2016) 1--25}.

\bibitem{Jevicki:2016ito}
A.~Jevicki and K.~Suzuki, \emph{{Bi-Local Holography in the SYK Model:
  Perturbations}}, \href{https://doi.org/10.1007/JHEP11(2016)046}{\emph{JHEP}
  {\bfseries 11} (2016) 046},
  [\href{https://arxiv.org/abs/1608.07567}{{\ttfamily 1608.07567}}].

\bibitem{Jevicki:2016bwu}
A.~Jevicki, K.~Suzuki and J.~Yoon, \emph{{Bi-Local Holography in the SYK
  Model}}, \href{https://doi.org/10.1007/JHEP07(2016)007}{\emph{JHEP}
  {\bfseries 07} (2016) 007},
  [\href{https://arxiv.org/abs/1603.06246}{{\ttfamily 1603.06246}}].

\bibitem{garcia2016}
A.~García-García and J.~Verbaarschot, \emph{{Spectral and thermodynamic
  properties of the Sachdev-Ye-Kitaev model}},
  \href{https://doi.org/10.1103/PhysRevD.94.126010}{\emph{Phys. Rev.}
  {\bfseries D94} (2016) 126010},
  [\href{https://arxiv.org/abs/1610.03816}{{\ttfamily 1610.03816}}].

\bibitem{witten2016}
E.~Witten, \emph{{An SYK-Like Model Without Disorder}},
  \href{https://arxiv.org/abs/1610.09758}{{\ttfamily 1610.09758}}.

\bibitem{Das:2017pif}
S.~R. Das, A.~Jevicki and K.~Suzuki, \emph{{Three Dimensional View of the
  SYK/AdS Duality}}, \href{https://doi.org/10.1007/JHEP09(2017)017}{\emph{JHEP}
  {\bfseries 09} (2017) 017},
  [\href{https://arxiv.org/abs/1704.07208}{{\ttfamily 1704.07208}}].

\bibitem{Das:2017wae}
S.~R. Das, A.~Ghosh, A.~Jevicki and K.~Suzuki, \emph{{Space-Time in the SYK
  Model}},  \href{https://arxiv.org/abs/1712.02725}{{\ttfamily 1712.02725}}.

\bibitem{cotler:2017jue}
J.~Cotler, N.~Hunter-Jones, J.~Liu and B.~Yoshida, \emph{{Chaos, Complexity,
  and Random Matrices}},
  \href{https://doi.org/10.1007/JHEP11(2017)048}{\emph{JHEP} {\bfseries 11}
  (2017) 048}, [\href{https://arxiv.org/abs/1706.05400}{{\ttfamily
  1706.05400}}].

\bibitem{cotler2016}
J.~Cotler, G.~Gur-Ari, M.~Hanada, J.~Polchinski, P.~Saad, S.~Shenker et~al.,
  \emph{{Black Holes and Random Matrices}},
  \href{https://doi.org/10.1007/JHEP05(2017)118}{\emph{JHEP} {\bfseries 05}
  (2017) 118}, [\href{https://arxiv.org/abs/1611.04650}{{\ttfamily
  1611.04650}}].

\bibitem{Krishnan:2016bvg}
C.~Krishnan, S.~Sanyal and P.~N. Bala~Subramanian, \emph{{Quantum Chaos and
  Holographic Tensor Models}},
  \href{https://doi.org/10.1007/JHEP03(2017)056}{\emph{JHEP} {\bfseries 03}
  (2017) 056}, [\href{https://arxiv.org/abs/1612.06330}{{\ttfamily
  1612.06330}}].

\bibitem{Turiaci:2017zwd}
G.~Turiaci and H.~Verlinde, \emph{{Towards a 2d QFT Analog of the SYK Model}},
  \href{https://doi.org/10.1007/JHEP10(2017)167}{\emph{JHEP} {\bfseries 10}
  (2017) 167}, [\href{https://arxiv.org/abs/1701.00528}{{\ttfamily
  1701.00528}}].

\bibitem{Klebanov:2016xxf}
I.~R. Klebanov and G.~Tarnopolsky, \emph{{Uncolored random tensors, melon
  diagrams, and the Sachdev-Ye-Kitaev models}},
  \href{https://doi.org/10.1103/PhysRevD.95.046004}{\emph{Phys. Rev.}
  {\bfseries D95} (2017) 046004},
  [\href{https://arxiv.org/abs/1611.08915}{{\ttfamily 1611.08915}}].

\bibitem{Stanford:2017thb}
D.~Stanford and E.~Witten, \emph{{Fermionic Localization of the Schwarzian
  Theory}}, \href{https://doi.org/10.1007/JHEP10(2017)008}{\emph{JHEP}
  {\bfseries 10} (2017) 008},
  [\href{https://arxiv.org/abs/1703.04612}{{\ttfamily 1703.04612}}].

\bibitem{gross2017}
D.~J. Gross and V.~Rosenhaus, \emph{{A Generalization of Sachdev-Ye-Kitaev}},
  \href{https://doi.org/10.1007/JHEP02(2017)093}{\emph{JHEP} {\bfseries 02}
  (2017) 093}, [\href{https://arxiv.org/abs/1610.01569}{{\ttfamily
  1610.01569}}].

\bibitem{Gross:2017aos}
D.~J. Gross and V.~Rosenhaus, \emph{{All point correlation functions in SYK}},
  \href{https://arxiv.org/abs/1710.08113}{{\ttfamily 1710.08113}}.

\bibitem{Bagrets:2017pwq}
D.~Bagrets, A.~Altland and A.~Kamenev, \emph{{Power-law out of time order
  correlation functions in the SYK model}},
  \href{https://doi.org/10.1016/j.nuclphysb.2017.06.012}{\emph{Nucl. Phys.}
  {\bfseries B921} (2017) 727--752},
  [\href{https://arxiv.org/abs/1702.08902}{{\ttfamily 1702.08902}}].

\bibitem{Cai:2017vyk}
W.~Cai, X.-H. Ge and G.-H. Yang, \emph{{Diffusion in higher dimensional SYK
  model with complex fermions}},
  \href{https://doi.org/10.1007/JHEP01(2018)076}{\emph{JHEP} {\bfseries 01}
  (2018) 076}, [\href{https://arxiv.org/abs/1711.07903}{{\ttfamily
  1711.07903}}].

\bibitem{maldacena2015}
J.~Maldacena, S.~Shenker and D.~Stanford, \emph{{A bound on chaos}},
  \href{https://doi.org/10.1007/JHEP08(2016)106}{\emph{JHEP} {\bfseries 08}
  (2016) 106}, [\href{https://arxiv.org/abs/1503.01409}{{\ttfamily
  1503.01409}}].

\bibitem{you2016}
Y.-Z. You, A.~Ludwig and C.~Xu, \emph{Sachdev-ye-kitaev model and
  thermalization on the boundary of many-body localized fermionic
  symmetry-protected topological states},
  \href{https://doi.org/10.1103/PhysRevB.95.115150}{\emph{Phys. Rev. B}
  {\bfseries 95} (Mar, 2017) 115150}.

\bibitem{Bagrets:2016cdf}
D.~Bagrets, A.~Altland and A.~Kamenev, \emph{{Sachdev–Ye–Kitaev model as
  Liouville quantum mechanics}},
  \href{https://doi.org/10.1016/j.nuclphysb.2016.08.002}{\emph{Nucl. Phys.}
  {\bfseries B911} (2016) 191--205},
  [\href{https://arxiv.org/abs/1607.00694}{{\ttfamily 1607.00694}}].

\bibitem{altland2017}
A.~Altland and D.~Bagrets, \emph{{Quantum ergodicity in the SYK model}},
  \href{https://arxiv.org/abs/1712.05073}{{\ttfamily 1712.05073}}.

\bibitem{verbaarschot1984}
J.~Verbaarschot and M.~Zirnbauer, \emph{Replica variables, loop expansion, and
  spectral rigidity of random-matrix ensembles},
  \href{https://doi.org/10.1016/0003-4916(84)90240-9}{\emph{Ann. Phys. (N.Y.)}
  {\bfseries 158} (1984) 78 -- 119}.

\bibitem{mon1975}
K.~K. Mon and J.~B. French, \emph{{Statistical Properties of Many Particle
  Spectra}}, \href{https://doi.org/10.1016/0003-4916(75)90045-7}{\emph{Annals.
  Phys.} {\bfseries 95} (1975) 90--111}.

\bibitem{Liu:2016rdi}
Y.~Liu, M.~A. Nowak and I.~Zahed, \emph{{Disorder in the Sachdev-Yee-Kitaev
  Model}}, \href{https://doi.org/10.1016/j.physletb.2017.08.054}{\emph{Phys.
  Lett.} {\bfseries B773} (2017) 647--653},
  [\href{https://arxiv.org/abs/1612.05233}{{\ttfamily 1612.05233}}].

\bibitem{erdos2014}
L.~{Erd{\H o}s} and D.~{Schr{\"o}der}, \emph{{Phase Transition in the Density
  of States of Quantum Spin Glasses}},
  \href{https://doi.org/10.1007/s11040-014-9164-3}{\emph{Mathematical Physics,
  Analysis and Geometry} {\bfseries 17} (Dec., 2014) 9164},
  [\href{https://arxiv.org/abs/1407.1552}{{\ttfamily 1407.1552}}].

\bibitem{Feng:2018zsx}
R.~Feng, G.~Tian and D.~Wei, \emph{{Spectrum of SYK model}},
  \href{https://arxiv.org/abs/1801.10073}{{\ttfamily 1801.10073}}.

\bibitem{fu2017}
W.~Fu, D.~Gaiotto, J.~Maldacena and S.~Sachdev, \emph{{Supersymmetric
  Sachdev-Ye-Kitaev models}}, \href{https://doi.org/10.1103/PhysRevD.95.069904,
  10.1103/PhysRevD.95.026009}{\emph{Phys. Rev.} {\bfseries D95} (2017) 026009},
  [\href{https://arxiv.org/abs/1610.08917}{{\ttfamily 1610.08917}}].

\bibitem{li2017}
T.~Li, J.~Liu, Y.~Xin and Y.~Zhou, \emph{{Supersymmetric SYK model and random
  matrix theory}}, \href{https://doi.org/10.1007/JHEP06(2017)111}{\emph{JHEP}
  {\bfseries 06} (2017) 111},
  [\href{https://arxiv.org/abs/1702.01738}{{\ttfamily 1702.01738}}].

\bibitem{kanazawa2017}
T.~Kanazawa and T.~Wettig, \emph{{Complete random matrix classification of SYK
  models with $\mathcal{N}=0$, $1$ and $2$ supersymmetry}},
  \href{https://doi.org/10.1007/JHEP09(2017)050}{\emph{JHEP} {\bfseries 09}
  (2017) 050}, [\href{https://arxiv.org/abs/1706.03044}{{\ttfamily
  1706.03044}}].

\bibitem{cumulants}
E.~A. Cornish and R.~A. Fisher, \emph{Moments and cumulants in the
  specification of distributions}, {\emph{Revue de l'Institut International de
  Statistique / Review of the International Statistical Institute} {\bfseries
  5} (1938) 307--320}.

\bibitem{flajolet2000}
P.~Flajolet and M.~Noy, \emph{Analytic Combinatorics of Chord Diagrams}.
\newblock Springer Berlin Heidelberg, Berlin, Heidelberg, 2000,
  \href{https://doi.org/10.1007/978-3-662-04166-6\_17}{10.1007/978-3-662-04166-6\_17}.

\bibitem{Berkooz:2018qkz}
M.~Berkooz, P.~Narayan and J.~Simon, \emph{{Chord diagrams, exact correlators
  in spin glasses and black hole bulk reconstruction}},
  \href{https://arxiv.org/abs/1806.04380}{{\ttfamily 1806.04380}}.

\bibitem{oeis}
N.~Sloane.
\newblock {The On-Line Encuclopedia of Integer Sequences, \url{
  http://oeis.org}, A036778}, 1964.

\bibitem{mehta2004}
M.~Mehta, \emph{Random matrices}.
\newblock Academic press, 2004.

\bibitem{verbaarschot:1994gr}
J.~J.~M. Verbaarschot, \emph{{Spectral sum rules and Selberg's integral
  formula}}, \href{https://doi.org/10.1016/0370-2693(94)90784-6}{\emph{Phys.
  Lett.} {\bfseries B329} (1994) 351--357},
  [\href{https://arxiv.org/abs/hep-th/9402008}{{\ttfamily hep-th/9402008}}].

\end{thebibliography}\endgroup

\end{document}